\documentclass[12pt,preprint]{aastex}

\usepackage{natbib}
\usepackage{graphicx}
\bibpunct{(}{)}{,}{a}{}{;}

\newcommand{\lapprox}{{_<}\atop{^\sim}}
\newcommand{\amin}{$^{\prime}$}
\newcommand{\asec}{$^{\prime \prime}$}
\newcommand{\adeg}{$^{\circ}$}

\slugcomment{Herschel WISH program}

\shorttitle{The Herschel WISH program}
\shortauthors{van Dishoeck et al.}

\begin{document}


\title{Water in Star-Forming Regions with the Herschel Space Observatory (WISH):\\ Overview of key program and first results
    }


\author{
E.F.~van~Dishoeck\altaffilmark{1,2},
L.E.~Kristensen\altaffilmark{1},
A.O.~Benz\altaffilmark{3},
E.A.~Bergin\altaffilmark{4},
P.~Caselli\altaffilmark{5,6},
J.~Cernicharo\altaffilmark{7},
F.~Herpin\altaffilmark{8},
M.R.~Hogerheijde\altaffilmark{1},
D.~Johnstone\altaffilmark{9,10},
R.~Liseau\altaffilmark{11},
B.~Nisini\altaffilmark{12},
R.~Shipman\altaffilmark{13},
M.~Tafalla\altaffilmark{14},
F.~van der Tak\altaffilmark{13,15},
F.~Wyrowski\altaffilmark{16},
Y.~Aikawa\altaffilmark{17},
R.~Bachiller\altaffilmark{14},
A.~Baudry\altaffilmark{8},
M.~Benedettini\altaffilmark{18},
P.~Bjerkeli\altaffilmark{11},
G.A.~Blake\altaffilmark{19},
S.~Bontemps\altaffilmark{8},
J.~Braine\altaffilmark{8},
C.~Brinch\altaffilmark{1},
S.~Bruderer\altaffilmark{3},
L.~Chavarr{\'{\i}}a\altaffilmark{8},
C.~Codella\altaffilmark{6},
F.~Daniel\altaffilmark{7},
Th.~de~Graauw\altaffilmark{13},
E.~Deul\altaffilmark{1},
A.M.~di~Giorgio\altaffilmark{18},
C.~Dominik\altaffilmark{20,21},
S.D.~Doty\altaffilmark{22},
M.L.~Dubernet\altaffilmark{23,24},
P.~Encrenaz\altaffilmark{25},
H.~Feuchtgruber\altaffilmark{2},
M.~Fich\altaffilmark{26},
W.~Frieswijk\altaffilmark{13},
A.~Fuente\altaffilmark{27},
T.~Giannini\altaffilmark{12},
J.R.~Goicoechea\altaffilmark{7},
F.P.~Helmich\altaffilmark{13},
T.~Jacq\altaffilmark{8},
J.K.~J{\o}rgensen\altaffilmark{28},
A.~Karska\altaffilmark{2},
M.J.~Kaufman\altaffilmark{29},
E.~Keto\altaffilmark{30},
B.~Larsson\altaffilmark{31},
B.~Lefloch\altaffilmark{32},
D.~Lis\altaffilmark{33},
M.~Marseille\altaffilmark{13},
C.~M$^{\rm c}$Coey\altaffilmark{26,34},
G.~Melnick\altaffilmark{30},
D.~Neufeld\altaffilmark{35},
M.~Olberg\altaffilmark{11},
L.~Pagani\altaffilmark{25},
O.~Pani{\'c}\altaffilmark{36},
B. Parise\altaffilmark{16},
J.C.~Pearson\altaffilmark{37},
R.~Plume\altaffilmark{38},
C.~Risacher\altaffilmark{13},
D.~Salter\altaffilmark{1},
J.~Santiago-Garc\'{i}a\altaffilmark{39},
P.~Saraceno\altaffilmark{18},
P.~St{\"a}uber\altaffilmark{3},
T.A.~van~Kempen\altaffilmark{1},
R.~Visser\altaffilmark{1},
S.~Viti\altaffilmark{40},
M.~Walmsley\altaffilmark{6},
S.F.~Wampfler\altaffilmark{3},
U.A.~Y{\i}ld{\i}z\altaffilmark{1}
}

\altaffiltext{1}{Leiden Observatory, Leiden University, PO Box 9513, 2300 RA Leiden, The Netherlands}
\altaffiltext{2}{Max Planck Institut f\"{u}r Extraterrestrische Physik, Giessenbachstrasse 1, 85748 Garching, Germany}
\altaffiltext{3}{Institute of Astronomy, ETH Zurich, 8093 Zurich, Switzerland}
\altaffiltext{4}{Department of Astronomy, University of Michigan, 500 Church Street, Ann Arbor, MI 48109-1042, USA}
\altaffiltext{5}{School of Physics and Astronomy, University of Leeds, Leeds LS2 9JT, UK}
\altaffiltext{6}{INAF - Osservatorio Astrofisico di Arcetri, Largo E. Fermi 5, 50125 Firenze, Italy}
\altaffiltext{7}{Centro de Astrobiolog{\'i}a, Departamento de Astrof{\'i}sica, CSIC-INTA, Carretera de Ajalvir, Km 4, Torrej{\'o}n de Ardoz. 28850, Madrid, Spain}
\altaffiltext{8}{Universit\'{e} de Bordeaux, Laboratoire d'Astrophysique de Bordeaux, France; CNRS/INSU, UMR 5804, Floirac, France}
\altaffiltext{9}{National Research Council Canada, Herzberg Institute of Astrophysics, 5071 West Saanich Road, Victoria, BC V9E 2E7, Canada}
\altaffiltext{10}{Department of Physics and Astronomy, University of Victoria, Victoria, BC V8P 1A1, Canada}
\altaffiltext{11}{Department of Radio and Space Science, Chalmers University of Technology, Onsala Space Observatory, 439 92 Onsala, Sweden}
\altaffiltext{12}{INAF - Osservatorio Astronomico di Roma, 00040 Monte Porzio catone, Italy}
\altaffiltext{13}{SRON Netherlands Institute for Space Research, PO Box 800, 9700 AV, Groningen, The Netherlands}
\altaffiltext{14}{Observatorio Astron\'{o}mico Nacional (IGN), Calle Alfonso XII,3. 28014, Madrid, Spain}
\altaffiltext{15}{Kapteyn Astronomical Institute, University of Groningen, PO Box 800, 9700 AV, Groningen, The Netherlands}
\altaffiltext{16}{Max-Planck-Institut f\"{u}r Radioastronomie, Auf dem H\"{u}gel 69, 53121 Bonn, Germany}
\altaffiltext{17}{Department of Earth and Planetary Sciences, Kobe University, Nada, Kobe 657-8501, Japan}
\altaffiltext{18}{INAF - Instituto di Fisica dello Spazio Interplanetario, Area di Ricerca di Tor Vergata, via Fosso del Cavaliere 100, 00133 Roma, Italy}
\altaffiltext{19}{California Institute of Technology, Division of Geological and Planetary Sciences, MS 150-21, Pasadena, CA 91125, USA}
\altaffiltext{20}{Astronomical Institute Anton Pannekoek, University of Amsterdam, Kruislaan 403, 1098 SJ Amsterdam, The Netherlands}
\altaffiltext{21}{Department of Astrophysics/IMAPP, Radboud University Nijmegen, P.O. Box 9010, 6500 GL Nijmegen, The Netherlands}
\altaffiltext{22}{Department of Physics and Astronomy, Denison University, Granville, OH, 43023, USA}
\altaffiltext{23}{Universit\'{e} Pierre et Marie Curie, LPMAA UMR CNRS 7092, Case 76, 4 place Jussieu, 75252 Paris Cedex 05, France}
\altaffiltext{24}{Observatoire de Paris-Meudon, LUTH UMR CNRS 8102, 5 place Jules Janssen, 92195 Meudon Cedex, France}
\altaffiltext{25}{LERMA and UMR 8112 du CNRS, Observatoire de Paris, 61 Av. de l'Observatoire, 75014 Paris, France}
\altaffiltext{26}{University of Waterloo, Department of Physics and Astronomy, Waterloo, Ontario, Canada}
\altaffiltext{27}{Observatorio Astron\'{o}mico Nacional, Apartado 112, 28803 Alcal\'{a} de Henares, Spain}
\altaffiltext{28}{Centre for Star and Planet Formation, Natural History Museum of Denmark, University of Copenhagen, {\O}ster Voldgade 5-7, DK-1350 Copenhagen K, Denmark}
\altaffiltext{29}{Department of Physics and Astronomy, San Jose State University, One Washington Square, San Jose, CA 95192, USA}
\altaffiltext{30}{Harvard-Smithsonian Center for Astrophysics, 60 Garden Street, MS 42, Cambridge, MA 02138, USA}
\altaffiltext{31}{Department of Astronomy, Stockholm University, AlbaNova, 106 91 Stockholm, Sweden}
\altaffiltext{32}{Laboratoire d'Astrophysique de Grenoble, CNRS/Universit\'{e} Joseph Fourier (UMR5571) BP 53, F-38041 Grenoble cedex 9, France}
\altaffiltext{33}{California Institute of Technology, Cahill Center for Astronomy and Astrophysics, MS 301-17, Pasadena, CA 91125, USA}
\altaffiltext{34}{University of Western Ontario, Department of Physics \& Astronomy, London, Ontario, Canada N6A 3K7}
\altaffiltext{35}{Department of Physics and Astronomy, Johns Hopkins University, 3400 North Charles Street, Baltimore, MD 21218, USA}
\altaffiltext{36}{European Southern Observatory, Karl-Schwarzschild-Str. 2, 85748 Garching, Germany}
\altaffiltext{37}{Jet Propulsion Laboratory, California Institute of Technology, Pasadena, CA 91109, USA}
\altaffiltext{38}{Department of Physics and Astronomy, University of Calgary, Calgary, T2N 1N4, AB, Canada}
\altaffiltext{39}{Instituto de RadioAstronom\'{i}a Milim\'{e}trica, Avenida Divina Pastora, 7, N\'{u}cleo Central E 18012 Granada, Spain}
\altaffiltext{40}{Department of Physics and Astronomy, University College London, Gower Street, London WC1E6BT, UK}

\email{ewine@strw.leidenuniv.nl}

\begin{abstract}
  `Water In Star-forming regions with Herschel' (WISH) is a key
  program on the {\it Herschel Space Observatory} designed to probe
  the physical and chemical structure of young stellar objects using
  water and related molecules and to follow the water abundance from
  collapsing clouds to planet-forming disks. About 80 sources are
  targeted covering a wide range of luminosities --from low ($<1$
  L$_\odot$) to high ($>10^5$ L$_\odot$)-- and a wide range of
  evolutionary stages --from cold pre-stellar cores to warm
  protostellar envelopes and outflows to disks around young
  stars. Both the HIFI and PACS instruments are used to observe a
  variety of lines of H$_2$O, H$_2$$^{18}$O and chemically related
  species at the source position and in small maps around the
  protostars and selected outflow positions.  In addition,
  high-frequency lines of CO, $^{13}$CO and C$^{18}$O are obtained
  with {\it Herschel} and are complemented by ground-based
  observations of dust continuum, HDO, CO and its isotopologues and
  other molecules to ensure a self-consistent data set for analysis.
  An overview of the scientific motivation and observational strategy
  of the program is given together with the modeling approach and
  analysis tools that have been developed. Initial science results are
  presented. These include a lack of water in cold gas at abundances
  that are lower than most predictions, strong water emission
  from shocks in protostellar environments, the importance of UV
  radiation in heating the gas along outflow walls across the full
  range of luminosities, and surprisingly widespread detection of the
  chemically related hydrides OH$^+$ and H$_2$O$^+$ in outflows and
  foreground gas.  Quantitative estimates of the energy budget
  indicate that H$_2$O is generally not the dominant coolant in the
  warm dense gas associated with protostars. Very deep limits on the
  cold gaseous water reservoir in the outer regions of protoplanetary
  disks are obtained which have profound implications for our
  understanding of grain growth and mixing in disks.

\end{abstract}


\section{Introduction} 
\label{sect:intro}

As interstellar clouds collapse to form new stars, part of the gas and
dust are transported from the infalling envelope to the rotating disk
out of which new planetary systems may form \citep{Shu87}. Water has a
pivotal role in these protostellar and protoplanetary environments
\citep[see reviews by][]{Cernicharo05,Melnick09}.
As a dominant form of oxygen, the most abundant element in the
universe after H and He, it controls the chemistry of many other
species, whether in gaseous or solid phase.  It is a unique diagnostic
of warm gas and energetic processes taking place during star
formation.  In cold regions, water is primarily in solid form, and its
presence as an ice may help the coagulation process that ultimately
produces planets. Asteroids and comets containing ice have likely
delivered most of the water to our oceans on Earth, where water is
directly associated with the emergence of life.  Water also
contributes to the energy balance as a gas coolant, allowing clouds to
collapse up to higher temperatures.
The distribution of water vapor and ice
during the entire star and planet formation process is therefore
a fundamental question relevant to our own origins.

The importance of water as a physical diagnostic stems from the orders
of magnitude variations in its gas phase abundance between warm and
cold regions
\citep[e.g.,][]{Cernicharo90,vanDishoeck96,Ceccarelli96,Harwit98,Ceccarelli99,Snell00,Nisini02,Maret02,Boonman03,vanderTak06}.
Thus, water acts like a `switch' that turns on whenever energy is
deposited in molecular clouds and elucidates key episodes in the
process of stellar birth, in particular when the system exchanges
matter and energy with its environment. This includes basic stages
like gravitational collapse, outflow injection, and stellar heating of
disks and envelopes. Its unique ability to act as a natural filter for
warm gas and probe cold gas in absorption make water a highly
complementary diagnostic to the commonly used CO molecule.

Studying water is also central to understanding the fundamental
processes of freeze-out, grain surface chemistry, and evaporation
\citep[e.g.,][]{Hollenbach09}.  Water is the dominant ice constituent
and can trap various molecules inside its matrix, including complex
organic ones \citep[e.g.,][]{Gibb04,Boogert08}. When and where water
evaporates back into the gas is therefore relevant for understanding
the wealth of organic molecules observed near protostars \citep[see
review by][]{Herbst09}. Young stars also emit copious UV radiation and
X-rays which affect the physical structure as well as the chemistry,
in particular that of hydrides like water \citep{Stauber06}. Moreover,
the level of deuteration of water provides an important record of the
temperature history of the cloud and the conditions during grain
surface formation, and, in comparison with cometary data, of its
evolution from interstellar clouds to solar system objects
\citep[e.g.,][]{Jacq90,Gensheimer96,Helmich96,Parise03}.

Finally, water plays an active role in the energy balance of dense gas
\citep[e.g.,][] {Goldsmith78,Neufeld93,Doty97}.  Because water has a
large dipole moment, its emission lines can be efficient coolants of
the gas, contributing significantly over the range of physical
conditions appropriate in star-forming regions. The large dipole
moment can also lead to heating through absorption of infrared
radiation followed by collisional de-excitation \citep{Ceccarelli96}.
It is therefore important to study the gaseous water emission and
absorption, and compare its cooling or heating efficiency
quantitatively with that of other species.

Interstellar water was detected more than 40 years ago through its 22
GHz maser emission towards Orion \citep{Cheung69}. While this line
remains a useful beacon of star-formation activity, its special
excitation and line formation conditions limit its usefulness as a
quantitative physical and chemical tool.  Because observations of
thermally excited water lines from Earth are limited, most information
on water has come from satellites. The {\it Submillimeter Wave
Astronomy Satellite} \citep[SWAS,][]{Melnick00} and the Odin satellite
\citep[e.g.,][]{Hjalmarson03,Bjerkeli09} observed only the 557 GHz
ground-state line of ortho-H$_2$O at spatial resolutions of $\sim
3.3'\times 4.5'$ and $126''$, respectively.  The 557 GHz line was
detected in just the brightest objects by these missions and other
lines of water (including those of para-H$_2$O) and related species
could not be observed.  The Short (SWS) and Long Wavelength
Spectrometers (LWS) on the {\it Infrared Space Observatory} (ISO)
covered a large number of pure rotational and vibration-rotation lines
providing important insight in the water excitation and
distribution, but with limited spectral and spatial resolution and
mapping capabilities \citep[see review by][]{vanDishoeck04}.  The {\it
Spitzer Space Telescope} has detected highly excited pure rotational
H$_2$O lines at mid-infrared wavelengths from shocks
\citep{Melnick08,Watson07} and from the inner few AU regions of
protoplanetary disks \citep{Salyk08,Carr08,Pontoppidan10}. Even
higher-lying vibration-rotation H$_2$O lines have been observed at
near-infrared wavelengths from the ground \citep{Salyk08}. However,
these data do not provide any information on the cooler water
reservoir where the bulk of the disk mass is.

\begin{figure}
\includegraphics[angle=0,width=0.6\textwidth]{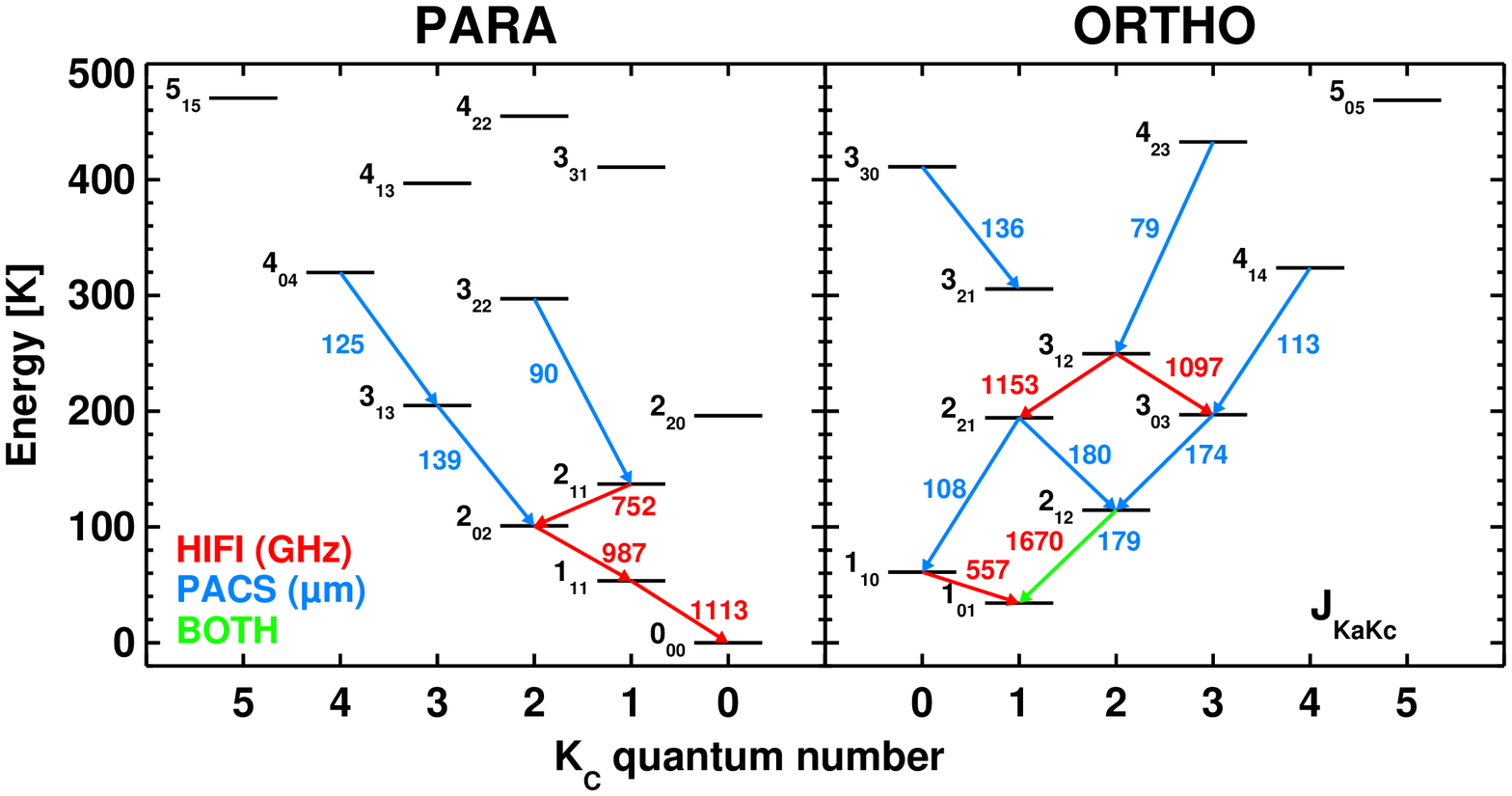}
\caption{Energy levels of ortho- and para-H$_2$O, with HIFI
  transitions (in GHz, red) and PACS transitions (in $\mu$m, blue)
  observed in WISH indicated.}
\label{fig:energy}
\end{figure}

The 3.5m {\it Herschel Space Observatory} with its suite of
instruments \citep{Pilbratt10}\footnote{Herschel is an ESA space
observatory with science instruments provided by European-led
Principal Investigator consortia and with important participation from
NASA} is particularly well suited to address the distribution of cold
and warm water in star- and planet-forming regions, building on
the pioneering results from previous missions. Its wavelength
coverage of 55--671 $\mu$m (0.45--5.4 THz; 15--180
cm$^{-1}$)\footnote{This paper follows the common usage of giving
frequencies in GHz for lines observed with HIFI and wavelengths in
$\mu$m for lines observed with PACS} includes both low- and
high-excitation lines of water, enabling a detailed analysis of its
excitation and abundance structure (Fig.~\ref{fig:energy}).  Compared
with the ISO-LWS beam ($\sim$80$''$), Herschel has a gain of a factor
of $\sim 8$ in spatial resolution at similar wavelengths, and up to
$10^4$ in spectral resolution.  Compared to observatories with high
spectral resolution heterodyne instruments, its diffraction limited
beam of 37$''$ at 557 GHz is a factor of 3--5 smaller than that of
SWAS or Odin, with a gain of $>10$ in sensitivity because of the
bigger dish and improved detector technology. The Heterodyne
Instrument for the Far-Infrared (HIFI; \citealt{deGraauw10}) has
spectral resolving powers $R=\lambda/\Delta \lambda>10^7$ up to 1900
GHz, allowing the kinematics of the water lines to be studied. The
Photoconducting Array Camera and Spectrometer (PACS;
\citealt{Poglitsch10}) $5\times 5$ pixel array provides instantaneous
spectral mapping capabilities at $R=1500-4000$ of important (backbone)
water lines in the 55--200 $\mu$m range. These enormous advances in
sensitivity, spatial and spectral resolution, and wavelength coverage
provide a unique opportunity to observe both cold and hot water in
space, with no other space mission with similar capabilities being
planned.

The goal of the `Water In Star-forming regions with Herschel' (WISH)
Key Program (KP) is to use gas-phase water as a physical and
chemical diagnostic and follow its abundance throughout the different
phases of star and planet formation.  A comprehensive set of water
observations is carried out with HIFI and PACS toward a large sample
of young stellar objects (YSOs), covering a wide range of masses and
luminosities -- from the lowest to the highest mass protostars-- and a
large range of evolutionary stages -- from the first stages
represented by the pre-stellar cores to the late stages represented by
the pre-main sequence stars surrounded only by disks.  Lines of
H$_2$O, H$_2^{18}$O, H$_2^{17}$O and the chemically related species O,
OH, and H$_3$O$^+$ are targeted\footnote{Both H$_2$O and `water' are
used to denote the main H$_2^{16}$O isotopologue throughout this
paper}.  In addition, a number of hydrides which are diagnostic of the
presence of X-rays and UV radiation are observed.  Selected
high-frequency lines of CO, $^{13}$CO, and C$^{18}$O as well as dust
continuum maps are obtained to constrain the physical structure of the
sources independent of the chemical effects. Together with the atomic
O and C$^+$ lines, these data also constrain the contributions from
the major coolants.  The Herschel data are complemented by
ground-based maps of longer wavelength continuum emission and lines of
HDO, CO, C and other molecules to ensure a self-consistent data set
for analysis. In terms of water, the WISH program is complementary to
the `CHEmical Survey of Star-forming regions' (CHESS;
\citealt{Ceccarelli10chess}) and `Herschel observations of
EXtraOrdinary Sources' (HEXOS; \citealt{Bergin10}) HIFI KPs which
survey the entire spectral range (including many lines of water) but
only for a limited number of sources and with smaller on-source
integration times per frequency setting. The `Dust, Ice and Gas in
Time' (DIGIT) (PI N.J.\ Evans) and `Herschel Orion Protostar Survey'
(HOPS) (PI T.\ Megeath) KPs complement WISH by carrying out full PACS
spectral scans for a larger sample of low-mass embedded YSOs
\citep[e.g.,][]{vanKempen10dkcha}. The `PRobing InterStellar Molecules
with Absorption line Studies' (PRISMAS) targets the water chemistry in
the diffuse interstellar gas \citep{Gerin10} whereas `Water and
related chemistry in the Solar System' (HssO) observes water in
planets and comets \citep{Hartogh10}.
Together, these Herschel data on water and related hydride lines will
provide a legacy for decades to come.

In the following sections, background information on the water
chemistry and excitation relevant for interpreting the data is
provided (\S 2). The observational details and organization of the
WISH program are subsequently described in \S 3, with the specific
goals and first results of the various subprograms presented in \S 4.
A discussion of the results across the various evolutionary stages and
luminosities is contained in \S 5 together with implications for the
water chemistry, with conclusions in \S 6.  Detailed information can
be found at the WISH KP website {\tt
  http://www.strw.leidenuniv.nl/WISH/}. This website includes links to
model results, modeling tools, complementary data and outreach
material.  These data and analysis tools will be useful not only for
{\it Herschel} but also for planning observations with the Atacama
Large Millimeter/submillimeter Array (ALMA) and future far-infrared
missions.

\section{H$_2$O chemistry and excitation}
\label{sect:background}

Star formation takes place in cold dense cores with temperatures
around 10 K and H$_2$ densities of at least $10^4$ cm$^{-3}$. Once
collapse starts and a protostellar object has formed, its central
luminosity heats the surrounding envelope to temperatures well above
100 K. Moreover, bipolar jets and winds interact with the envelope and
cloud, creating shocks in which the temperature is raised to several
thousand K. Cloud core rotation leads to a circumstellar disk around
the young star with densities in the midplane of $>10^{10}$ cm$^{-3}$.
With time, the envelope is dispersed by the action of the outflow,
leaving the pre-main sequence star with a disk only. Once accretion
onto the disk stops and the disk becomes less turbulent, the $\sim$0.1
$\mu$m grains from the collapsing cloud coagulate to larger particles
and settle to the midplane, eventually leading to planetesimals and
(proto)planets.  Thus, star-forming regions contain gas with a large
range of temperatures and densities, from 10 to 2000 K and $10^4$ to
$>10^{10}$ cm$^{-3}$, and with gas/dust ratios that may vary from the
canonical value of 100 to much larger or smaller values.  The water
chemistry responds to these different conditions.

\begin{figure}
\includegraphics[angle=-90,width=0.7\textwidth]{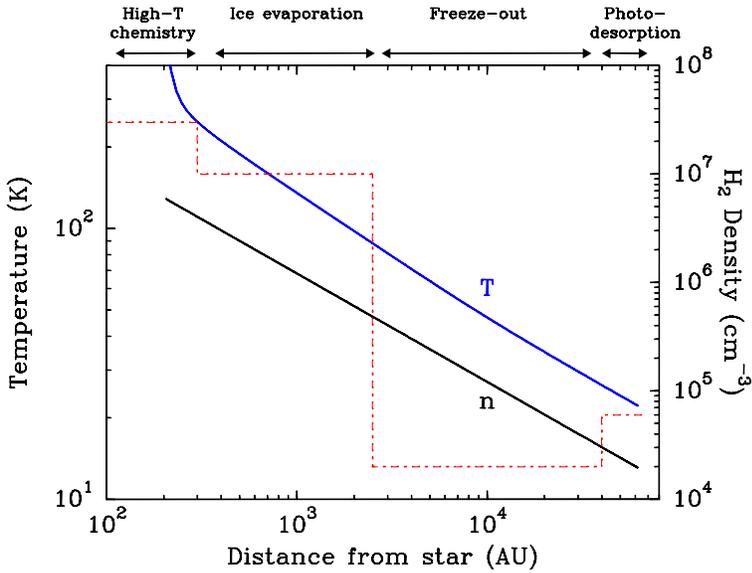}
\caption{Temperature (blue) and density (black) structure of a
  high-mass protostellar envelope. For illustrative purposes, a
  typical abundance profile is included (red dash-dotted line) on an
  arbitrary scale, with the relevant chemical processes in each regime
  indicated at the top of the figure.  The figure is based on the
  model of AFGL 2591 ($L=6\times 10^4$ L$_{\odot}$) presented by
  \citet{vanderTak99}.  For distances of $\sim$1 kpc, the typical Herschel
  beam of 20$''$ corresponds to 20000 AU, i.e., it encompasses the
  entire envelope. For low-mass protostars, the ice evaporation zone
  moves inward to radii typically less than 100 AU. }
\label{fig:nt}
\end{figure}

\begin{figure}
\includegraphics[angle=0,width=0.6\textwidth]{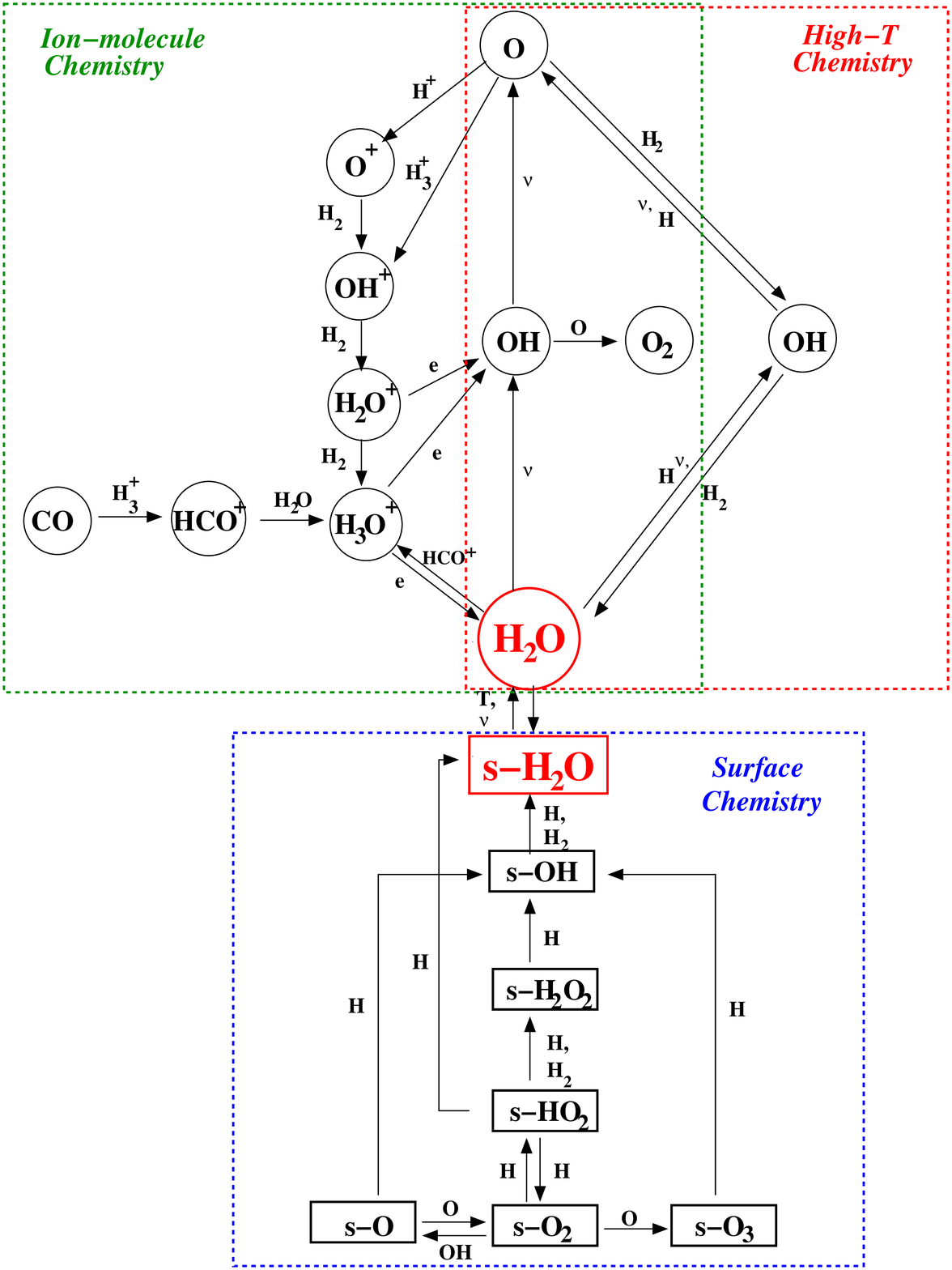}
\caption{Summary of the main gas-phase and solid-state chemical
  reactions leading to the formation and destruction of H$_2$O.  Three
  different chemical regimes can be distinguished: (i) ion-molecule
  chemistry, which dominates gas-phase chemistry at low $T$; (ii)
  high-temperature neutral-neutral chemistry; and (iii) solid state
  chemistry. The latter chemical scheme is based on the latest
  laboratory data by \citet{Ioppolo10o2}. s-$X$ denotes species $X$ on
  the ice surfaces.}
\label{fig:network}
\end{figure}

\subsection{Chemistry}
\label{sect:chemistry}

Figure \ref{fig:nt} contains an overview of the physical structure of
a typical protostellar envelope, with the main chemical regimes
indicated.  In standard gas-phase chemistry, H$_2$O forms through
ion-molecule reactions starting with O + H$_3^+$ and O$^+$ + H$_2$,
both leading to OH$^+$ \citep[e.g.,][]{Herbst73,Dalgarno76} (see
Fig.~\ref{fig:network}).  A series of rapid hydrogen abstraction
reactions with H$_2$ leads to H$_3$O$^+$, which can dissociatively
recombine to form H$_2$O and OH with branching ratios of $\sim$33\%
and 66\%, respectively \citep{Vejby97}. H$_2$O is destroyed by
ultraviolet photons and by reactions with C$^+$, H$_3^+$ and other
ions such as HCO$^+$. All key reactions in this network have been
measured or calculated at low temperatures
\citep[e.g.,][]{Klippenstein10}. The steady-state H$_2$O abundance
resulting from pure gas-phase chemistry is typically a few times
$10^{-7}$ with respect to H$_2$ at low $T$ and scales with the
cosmic-ray ionization rate $\zeta$ \citep[e.g.,][]{Umist07}.

In the last decades, this traditional view of the oxygen chemistry has
been radically changed, triggered by the very low abundances of O$_2$
in cold clouds found by SWAS \citep[e.g.,][] {Bergin00}.  It is now
realized that in cold and dense regions, H$_2$O is formed much more
efficiently on the grains through a series of reactions involving O
and H accreted from the gas, as predicted more than 50 years ago by
van de Hulst and quantified by \citet{Tielens82}. Surface science
laboratory experiments are only now starting to test these reactions
experimentally
\citep{Ioppolo08,Ioppolo10o2,Miyauchi08,Dulieu10,Romanzin10}. While
the details are not yet fully understood \citep[e.g.,][]{Cuppen10},
there is ample observational evidence from 3 $\mu$m absorption
spectroscopy of the solid H$_2$O band toward YSOs that the H$_2$O ice
abundances can be as high as $\sim 10^{-4}$, much higher than results
from freeze-out of gas-phase H$_2$O
\citep[e.g.,][]{Whittet88,Pontoppidan04}.  

Water ice formation starts
at a threshold extinction of $A_V\approx 5$ mag, i.e., well before the
cloud collapses \citep{Whittet03}. A small fraction of the H$_2$O ice
formed by surface reactions can be desorbed back into the gas phase by
a variety of non-thermal processes including cosmic-ray induced
desorption and photodesorption \citep[see recent overview
by][]{Hollenbach09}. The efficiency of the latter process has been
quantified by laboratory experiments to be $\sim 10^{-3}$ per incident
UV photon \citep{Westley95,Oberg09h2o}.

Near protostars, the grain temperature rises above $\sim 100$~K and
all the H$_2$O ice thermally desorbs on very short timescales
\citep{Fraser01}, leading to initial gas-phase abundances of
H$_2$O as high as the original ice abundances of $\sim
10^{-4}$. Other molecules trapped in the water ice matrix evaporate at
the same time in these so-called `hot core' or `hot corino' (in the
case of low-mass YSOs) regions. Thus, constraining the H$_2$O
evaporation zone is also critical for the interpretation of complex
organic molecules in hot cores.

At even higher temperatures, above 230~K, the gas-phase reaction O +
H$_2$ $\to$ OH + H, which is endoergic by $\sim$2000 K, becomes
significant \citep{Elitzur78,Charnley97}. OH subsequently reacts with
H$_2$ to form H$_2$O, a reaction which is exothermic but has an energy
barrier of $\sim$2100 K \citep{Atkinson04}.  This route drives all the
available gas-phase oxygen into H$_2$O leading to an abundance of
$\sim 3\times 10^{-4}$ at high temperatures, unless the H/H$_2$ ratio
in the gas is so high that the back-reactions become important.  Such
hot gas can be found in the hot cores close to the protostars
themselves and in the more extended shocked gas associated with the
outflows. Given the range of temperatures and densities in
protostellar (Fig.~\ref{fig:nt}) and protoplanetary environments, all
of the above chemical processes likely play a role, and the Herschel
data are needed to determine their relative importance.

H$_2$O has a unique deuteration pattern compared with other
molecules. If grain surface formation dominates, water is deuterated
in the ice at a level that may be orders of magnitude lower than that
of other species which participate more actively in the dense cold gas
chemistry \citep{Roberts03}.
Indeed, the deuteration fraction observed in high mass hot cores is typically
HDO/H$_2$O$<10^{-3}$ \citep[e.g.,][]{Gensheimer96,Helmich96}, although
higher values around $10^{-2}$ have recently been found for Orion
\citep{Persson07,Bergin10hexos} and for a low-mass hot core
\citep{Parise05hdo}.  All of these values are still much lower than,
for example, DCN/HCN, HDCO/H$_2$CO \citep[e.g.,][]{Loinard01} or
CH$_2$DOH/CH$_3$OH \citep{Parise04}.  In the solid phase, the
HDO/H$_2$O upper limits are less than $2\times 10^{-3}$
\citep{Dartois03,Parise03}. An important question is how similar the
protostellar HDO/H$_2$O ratios are to those observed in comets and in
water in our oceans on Earth (about $2\times 10^{-4}$), since this has
major implications for the delivery mechanisms of water to our planet
\citep{Bockelee98,Raymond04}.  No specific HDO lines are targeted with
      {\it Herschel} within the WISH program, since several HDO lines
      can be observed from the ground \citep{Parise05model}. HIFI will
      provide the necessary data on H$_2$O itself for comparison with
      those HDO data.

\subsection{Excitation}
\label{sect:excitation}

H$_2$O is an asymmetric rotor with a highly irregular set of energy
levels characterized by quantum numbers $J_{K_A K_C}$. Because of the
nuclear spin statistics of the two hydrogen atoms, the H$_2$O energy
levels are grouped into ortho ($K_A+K_C=$odd) and para
($K_A+K_C=$even) ladders across which no transitions can occur except
through chemical reactions which exchange a H nucleus
(Fig.~\ref{fig:energy}).
The H$_2$O energy levels within each ladder are populated by a
combination of collisional and radiative processes. The most important
collision partners are ortho- and para-H$_2$, with electrons and H
only significant in specific regions and He contributing at a low
level. Accurate collisional rate coefficients are thus crucial to
interpret the {\it Herschel} data.  The bulk of these rates are
obtained from theoretical quantum chemistry calculations which involve
two steps. First, a multi-dimensional potential energy surface
involving the colliders is computed.  Second, the dynamics on this
surface are investigated with molecular scattering calculations at a
range of collision energies. Early studies used simplifications such
as replacing H$_2$ with He \citep{Green93}, including only a limited
number of degrees of freedom of the colliding system, or using
approximations in the scattering calculations \citep{Phillips96}.  Recent 
calculations consider a 5D potential surface and compute
collisions with both para-H$_2$ and ortho-H$_2$ separately up to high
temperatures with the close coupling method
\citep{Dubernet09,Daniel10}.  
This 5D potential is based on  the full 9D potential surface 
computed by \citet{Valiron08}. 
The cross sections for collisions of
H$_2$O with $o-$H$_2$ are found to be significantly larger than those
with $p$-H$_2$ for low $J$, so that explicit treatment of the $o/p$
H$_2$ ratio is important.

Direct comparison of absolute values of theoretical state-to-state
cross sections with experimental data is not possible, but the
accuracy of the 9D potential surface has been confirmed implicitly
through comparison with other data sets, including differential
scattering experiments of H$_2$O with H$_2$ \citep{Yang10b}.  An
indication of the accuracy of the calculations can also be obtained
from recent pressure broadening experiments \citep{Dick10}. The high
temperature ($>80$ K) data agree with theory within 30\% or better,
but the low temperature experimental rate coefficients are an order of
magnitude smaller.  This discrepancy has recently been understood by
realizing that the classical impact approximation used in the analysis
of the pressure broadening data does not hold at very low temperatures
\citep{Wiesenfeld10}. The computed rate coefficients at low $T$ by
\citet{Dubernet06} should therefore be valid.

Because of the large rotation constants of water, the energy spacing
between the lower levels is much larger than that of a heavy molecule
like CO. Thus, the water transitions couple much more efficiently with
far-infrared radiation from warm dust, which can pump higher energy
levels \citep[e.g.,][]{Takahashi83}. This far-infrared continuum also
plays a role in the line formation (e.g., whether the line occurs in
absorption or emission) and thus provides constraints on source
geometry.  Moreover, the continuum can become optically thick at the
highest frequencies, providing an effective `screen' for looking to
different depths in the YSO environment \citep{Poelman07,vanKempen08}.

The high transition frequencies combined with the large dipole moment
of H$_2$O (1.85 Debye) also lead to high line opacities. For example,
for the $o-$H$_2$O 557 GHz ground state transition, the line center
optical depth is $\tau_0=2.0\times 10^{-13} N_\ell(o-{\rm H_2O})
\Delta V $, where $N_\ell$ is the column density in the lower level in
cm$^{-2}$, which is to first approximation equal to the total
$o-$H$_2$O column density, and $\Delta V$ is the FHWM of the line in
km s$^{-1}$ \citep{Plume04}. Thus, even for H$_2$O abundances as low
as $10^{-9}$ in typical clouds with $N$(H$_2$)$\geq 10^{22}$
cm$^{-2}$, the water lines are optically thick at line center.

Because of the high optically depths, it is difficult to extract
reliable water abundances from lines of the main
isotopologue. However, because the excitation is subthermal (critical
densities are typically $10^8-10^9$ cm$^{-3}$), every photon that is
created will scatter and eventually escape the cloud. In this
so-called effectively thin limit, the water column density scales
linearly with integrated intensity of a ground-state transition and is
inversely proportional to density and the value of the collisional
rate coefficient \citep{Snell00,Schulz91}.

Many of the H$_2$O transitions exhibit population inversion, either
very weakly or more strongly, such as the famous $6_{16}-5_{23}$ maser
at 22 GHz widely observed in star-forming regions
\citep[e.g.,][]{Szymczak05}.  Thus, the H$_2$O molecule forms a
formidable challenge for radiative transfer codes, and the HIFI and
PACS spectra provide a unique reference data set of thermal lines
against which to test the basic assumptions in the maser models.

\subsection{Modeling tools}

The WISH team has developed a variety of tools important for the
Herschel data analysis, several of which are publicly
available to the community.  This includes a molecular line database
LAMDA\footnote{\tt http://www.strw.leidenuniv.nl/$\sim$moldata} for
excitation calculations of H$_2$O, OH, CO and other molecules using
various sets of collisional rate coefficients discussed above
\citep{Schoier05}.  
Simple 1D escape probability non-LTE radiative transfer programs such
as RADEX can be run either on-line or downloaded to run at the
researcher's own institute \citep{vanderTak07}. This and the more
sophisticated Monte-Carlo 1D radiative transfer code RATRAN are
publicly available\footnote{\tt
  http://www.sron.rug.nl/$\sim$vdtak/ratran/frames.html} and RADEX is
also included in the CASSIS set of analysis programs\footnote{\tt
  http://www.cesr.fr/$\sim$walters/web$_{-}$cassis/}. The RATRAN code
has been extensively tested in a code comparison campaign
\citep{vanZadelhoff02}.  Another 1D non-LTE code
used by the WISH team is MOLLIE \citep{Keto04}.  A 2D version
of RATRAN as well as a new flexible 3D line radiative transfer code
called LIME are available on a collaborative basis
\citep{Hogerheijde00,Brinch10}.  A 2D escape probability code has been
written by \citet{Poelman05} whereas a fast 3D code using a local
source approximation has been developed by \citet{Bruderer10mod}.


\begin{deluxetable}{l cc c rr l c r c}
\tablecolumns{10}
\tabletypesize{\scriptsize}
\tablenum{1}
\tablewidth{0pc}
\tablecaption{WISH Source List}
\tablehead{
	\colhead{} &  \multicolumn{2}{c}{Coordinates} & \colhead{} & \multicolumn{2}{c}{Properties} & \colhead{} & \colhead{} & \colhead{} & \colhead{} \\
\cline{2-3} \cline{5-6} \\
\colhead{} & \colhead{RA}   & \colhead{Dec}    & \colhead{$V_{\rm LSR}$} & \colhead{$L_{\rm bol}$}    & \colhead{$d$}   & \colhead{Outflow\tablenotemark{a}}    & \colhead{RD\tablenotemark{b}} & \colhead{Ref.} & \colhead{Notes} \\
\colhead{} & \colhead{(h m s)}   & \colhead{(\adeg\ \amin\ \asec)}    & \colhead{(km\,s$^{-1}$)} & \colhead{($L_\odot$)}  & \colhead{(pc)}   & \colhead{(\asec\ \asec)}  & \colhead{}  & \colhead{} & \colhead{} \\}
\startdata
\cutinhead{Prestellar cores}
L\,1544     & 05 04 17.2 & $+$25 10 42.8  &  $+7.3$ & \nodata & 140 &  &  &  &  \\
B\,68       & 17 22 38.2 & $-$23 49 54.0  &  $+3.4$ & \nodata & 125 &  &  &  &  \\
\cutinhead{Low-mass YSOs}
L\,1448-MM         & 03 25 38.9 & $+$30 44 05.4  & $+5.3$  & 11.6 & 250  &  R $(+30,\,-125.4)$ &   & 1,2,3 & \tablenotemark{c} \\
                   &            &                &         &      &      &  B $(-15,\,+28.8)$  &   &       &  \\
NGC\,1333 IRAS\,2  & 03 28 55.6 & $+$31 14 37.1  & $+7.7$  & 20.7 & 235  &  B $(-92,\,+28)$    & + & 4,2,3 &  \\
                   &            &                &         &      &      &  R $(+70,\,-15)$    &   &       &  \\
NGC\,1333 IRAS\,3A & 03 29 03.8 & $+$31 16 04.0  & $+8.5$  & 50   & 235  & B1 $(+20,\,-20)$    & + & 1,5,3 &  \\
                   &            &                &         &      &      & B2 $(+20,\,-50)$    & + &       &  \\
NGC\,1333 IRAS\,4A & 03 29 10.5 & $+$31 13 30.9  & $+7.2$  & 7.7  & 235  &  B $(-6,\,-19)$     &   & 1,2,3 &  \\
                   &            &                &         &      &      &  R $(+13,\,+25)$    &   &       &  \\
NGC\,1333 IRAS\,4B & 03 29 12.0 & $+$31 13 08.1  & $+7.4$  & 7.7  & 235  &                     &   & 1,2,3 &  \\
L\,1527            & 04 39 53.9 & $+$26 03 09.8  & $+5.9$  & 2    & 140  &  B $(+40,\,+10)$    &   & 1,2,6 &  \\
BHR\,71            & 12 01 36.3 & $-$65 08 53.0  & $-4.4$  & 10   & 200  &  B $(+40,\,-100)$   &   & 6,7,8 &  \\
                   &            &                &         &      &      &  R $(-40,\,+140)$   &   &       &  \\
L\,483 MM          & 18 17 29.9 & $-$04 39 39.5  & $+5.2$  & 9    & 200  &  B $(-40,\,0)$      &   & 1,2,6 &  \\
Ser SMM\,1         & 18 29 49.8 & $+$01 15 20.5  & $+8.5$  & 30   & 250\tablenotemark{l}  & B1 $(-15,\,+30)$    & + & 9,10,11 &  \\
Ser SMM\,4         & 18 29 56.6 & $+$01 13 15.1  & $+8.0$  & 5.0  & 250\tablenotemark{l}  &  R $(+30,\,-60)$    &   & 9,12,11 &  \\
Ser SMM\,3         & 18 29 59.2 & $+$01 14 00.3  & $+7.6$  & 5.9  & 250\tablenotemark{l}  &                     &   & 9,12,11 &  \\
L\,723 MM          & 19 17 53.7 & $+$19 12 20.0  & $+11.2$ & 3    & 300  &                     &   & 6,13 &  \\
B\,335             & 19 37 00.9 & $+$07 34 09.6  & $+8.4$  & 3    & 250  &  B $(+30,\,0)$      &   & 1,2,6 &  \\
L\,1157            & 20 39 06.3 & $+$68 02 15.8  & $+2.6$  & 6    & 325  & B2 $(+35,\,-95)$    &   & 1,2,6 &  \\
                   &            &                &         &      &      &  R $(-30,\,+140)$   &   &       &  \\
\tableline
L\,1489            & 04 04 43.0 & $+$26 18 57.0  & $+7.2$ &   3.7 & 140  &                     & + & 14,15,6 & \tablenotemark{d} \\
L\,1551 IRS\,5     & 04 31 34.1 & $+$18 08 05.0  & $+7.2$ &  28   & 140  &  B $(-255,\,-255)$  &   & 16,12,6 &  \\
                   &            &                &         &      &      &  R $(+150,\,+20)$   &   &       &  \\
TMR\,1\tablenotemark{m}             & 04 39 13.7 & $+$25 53 21.0  & $+6.3$ &   3.7 & 140  &                     &   & 4,12,6 &  \\
TMC\,1A\tablenotemark{m}            & 04 39 34.9 & $+$25 41 45.0  & $+6.6$ &   2.2 & 140  &                     &   & 12,6 &  \\
TMC\,1             & 04 41 12.4 & $+$25 46 36.0  & $+5.2$ &   0.7 & 140  &                     &   & 12,6 &  \\
HH\,46             & 08 25 43.9 & $-$51 00 36.0  & $+5.2$ &  12   & 450  &  B $(-10,\,0)$      &   & 17,18 &  \\
                   &            &                &         &      &      &  R $(-40,\,-20)$    &   &       &  \\
Ced110 IRS4        & 11 06 47.0 & $-$77 22 32.4  & $+3.5$ &   1   & 125  &                     &   & 19,1,20 &  \\
IRAS 12496/DK\,Cha & 12 53 17.2 & $-$77 07 10.6  & $+2.3$ &  50   & 200  &                     &   & 1,20 &  \\
IRAS 15398/B228    & 15 43 01.3 & $-$34 09 15.0  & $+5.1$ &   1   & 130  &                     &   & 1,20 &  \\
GSS 30 IRS1        & 16 26 21.4 & $-$24 23 04.0  & $+2.8$ &  25   & 125  &                     &   & 1,4,21 &  \\
Elias 29           & 16 27 09.4 & $-$24 37 19.6  & $+5.0$ &  36   & 125  &                     &   & 1,22,21 &  \\
Oph IRS 63         & 16 31 35.6 & $-$24 01 29.6  & $+2.8$ &   1.6 & 125  &                     &   & 1,22,21 &  \\
RNO 91             & 16 34 29.3 & $-$15 47 01.4  & $+5.0$ &  11   & 125  &                     &   & 23,1,21 &  \\
R CrA IRS 5        & 19 01 48.0 & $-$36 57 21.6  & $+5.7$ &   3   & 170  &                     &   & 24,1,20 &  \\ 
HH 100             & 19 01 49.1 & $-$36 58 16.0  & $+5.6$ &  14   & 125  &                     &   & 25,1,20 &  \\
\cutinhead{Intermediate-mass YSOs}
AFGL\,490         & 03 27 38.4 & $+$58 47 08.0 & $-$13.5 & 2000 & 1000 &                  &   & 26,27,28 &  \\
L1641 S3\,MMS1    & 05 39 55.9 & $-$07 30 28.0 & $+$5.3  &   70 &  465 &                  &   & 29,30 &  \\
NGC\, 2071        & 05 47 04.4 & $+$00 21 49.0 & $+$9.6  &  520 &  422 & P0 $(+32,\,+63)$ &   & 31,32,30 &  \\
                  &            &               &         &      &      & P5 $(-128,\,-97)$ &  &          &  \\
Vela IRS\,17      & 08 46 34.7 & $-$43 54 30.5 & $+$3.9  &  715 &  700 &                  &   & 33,34,35 &  \\
Vela IRS\,19      & 08 48 48.5 & $-$45 32 29.0 & $+$12.2 &  776 &  700 &                  &   & 33,35 &  \\
NGC\,7129 FIRS\,2 & 21 43 01.7 & $+$66 03 23.6 & $-$9.8  &  430 & 1250 & B $(+60,\,+60)$  & + & 36,37,38 &  \\
                  &            &               &         &      &      & R $(+60,\,-60)$  &   &          &  \\
\cutinhead{Outflow only}
HH\,211-mm     & 03 43 56.8 & $+$32 00 50  & $+9.2$ &  4\phantom{.0} & 250 & C $(0,\,0)$       &  &  &  \\
               &            &              &        &                &     & B2 $(+37,\,-15)$  &  &  &  \\
IRAS\,04166    & 04 19 42.6 & $+$27 13 38  & $+6.7$ & 0.4            & 140 & B $(+20,\,+35)$   &  &  &  \\
               &            &              &        &                &     & R $(-20,\,-35)$   &  &  &  \\
VLA-1 HH1/2    & 05 36 22.8 & $-$06 46 07  & $+8$   & 50\phantom{.0} & 450 & B $(+60,\,-80)$   &  &  &  \\
HH\,212 MM1    & 05 43 51.4 & $-$01 02 53  & $+1.6$ & 15\phantom{.0} & 460 & C $(0,\,0)$       &  &  &  \\
               &            &              &        &                &     & B $(-15,\,-35)$   &  &  &  \\
HH\,25\,MMS    & 05 46 07.3 & $-$00 13 30  & $+10$  &  6\phantom{.0} & 400 & C $(0,\,0)$       &  &  &  \\
               &            &              &        &                &     & SiO $(+36,\,-57)$ &  &  &  \\
HH\,111 VLA\,1 & 05 51 46.3 & $+$02 48 30  & $+8.5$ & 25\phantom{.0} & 460 & C $(0,\,0)$       &  &  &  \\
               &            &              &        &                &     & B1 $(-170,\,+21)$ &  &  &  \\
HH\,54B        & 12 55 50.3 & $-$76 56 23  & $+$2.4 &  1\phantom{.0} & 180 & C $(0,\,0)$       &  &  &  \\
VLA\,1623      & 16 26 26.4 & $-$24 24 30  & $+3.5$ &  1\phantom{.0} & 125 & B1 $(+30,\,-20)$  &  &  &  \\
               &            &              &        &                &     & R2 $(-65,\,+25)$  &  &  &  \\
IRAS\,16293    & 16 32 22.8 & $-$24 28 36  & $+4.5$ & 14\phantom{.0} & 125 & B $(+75,\,-60)$   &  &  &  \\
               &            &              &        &                &     & R $(+75,\,+45)$   &  &  &  \\
Ser S\,68\,N   & 18 29 47.5 & $+$01 16 51  & $+8.5$ &  4\phantom{.0} & 260 & B $(-12,\,+24)$   &  &  &  \\
               &            &              &        &                &     & C $(0,\,0)$       &  &  &  \\
Cep E\, MM	   & 23 03 13.1 & $+$61 42 26  & $-$11.0 &            75 & 730 & B $(-10,\,-20)$   &  &  &  \\
\cutinhead{High-mass YSOs}
G11.11$-$0.12-NH$_{\rm 3}$-P1   & 18 10 33.9    &  $-$19 21 48   &  +30.4   & \nodata     & 3600 &  &  & 39, 40 & \tablenotemark{e} \\
G11.11$-$0.12-SCUBA-P1          & 18 10 28.4    &  $-$19 22 29   &  +29.2   & \nodata     & 3600 &  &  & 40 \\
G28.34+0.06-NH$_{\rm 3}$-P3     & 18 42 46.4    &  $-$04 04 12   &  +80.2   & \nodata     & 4800 &  &  & 40 \\
G28.34+0.06-SCUBA-P2            & 18 42 52.4    &  $-$03 59 54   &  +78.5   & \nodata     & 4800 &  &  & 40 \\
\tableline
IRAS05358+3543      & 05 39 13.1    &   +35 45 50   & $-$17.6   & $6.3\,\times\,10^3$     & 1800 &  &   & 41 & \tablenotemark{f}  \\
IRAS16272$-$4837    & 16 30 58.7    & $-$48 43 55   & $-$46.2   & $2.4\,\times\,10^4$     & 3400 &  &   & 42  \\
NGC6334-I           & 17 20 53.3    & $-$35 47 00   & $-$4.5    & $1.7\,\times\,10^4$     & 1700 &  & + & 43  \\
W43-MM1             & 18 47 47.0    & $-$01 54 28   & +98.8     & $2.3\,\times\,10^4$     & 5500 &  &   & 44 \\
DR21(OH)            & 20 39 00.8    &   +42 22 48   & $-$4.5    & $1.7\,\times\,10^4$     & 1700 &  &   & 45, 46 \\
\tableline
W3-IRS5             & 02 25 40.6    &   +62 05 51   & $-$38.4   & $1.7\,\times\,10^5$     & 2200 &  & + & 47, 48, 49 & \tablenotemark{g} \\
IRAS18089$-$1732    & 18 11 51.5    & $-$17 31 29   & +33.8     & $3.2\,\times\,10^4$     & 3600 &  &   & 50 \\
W33A                & 18 14 39.5    & $-$17 52 00   & +37.5     & $1.0\,\times\,10^4$     & 4000 &  &   & 51  \\
IRAS18151$-$1208    & 18 17 58      & $-$12 07 27   & +32.0     & $2.0\,\times\,10^4$     & 2900 &  &   & 52 \\
AFGL2591            & 20 29 24.9    & +40 11 19.5   & $-$5.5    & $5.8\,\times\,10^4$     & 1700 &  & + & 53 \\
\tableline
G327$-$0.6          & 15 53 08.8    & $-$54 37 01   & $-$45.0   & $1.0\,\times\,10^5$     & 3000 &  &   & 54, 55 & \tablenotemark{h} \\
NGC6334-I(N)        & 17 20 55.2    & $-$35 45 04   & $-$7.7    & $1.1\,\times\,10^5$     & 1700 &  & + & 56 \\
G29.96$-$0.02       & 18 46 03.8    & $-$02 39 22   & +98.7     & $1.2\,\times\,10^5$     & 7400 &  &   & 57, 58 \\
G31.41+0.31         & 18 47 34.3    & $-$01 12 46   & +98.8     & $1.8\,\times\,10^5$     & 7900 &  &   & 59, 60, 61 \\
IRAS20126+4104      & 20 14 25.1    & $+$41 13 32   & $-$3.8    & $1.0\,\times\,10^4$     & 1700 &  &   & 50 \\
\tableline
G5.89$-$0.39        & 18 00 30.4    & $-$24 04 02   & +10.0     & $2.5\,\times\,10^4$     & 2000 &  &   & 62, 63 & \tablenotemark{i} \\
G10.47+0.03         & 18 08 38.2    & $-$19 51 50   & +67.0     & $1.1\,\times\,10^5$     & 5800 &  &   & 64, 65, 57 \\
G34.26+0.15         & 18 53 18.6    &   +01 14 58   & +57.2     & $2.8\,\times\,10^5$     & 3300 &  &   & 66, 65 \\
W51N-e1             & 19 23 43.8    &   +14 30 26   & +59.5     & $1 - 10\,\times\,10^5$  & 5500 &  &   & 62 \\
NGC7538-IRS1        & 23 13 45.3    &   +61 28 10   & $-$57.4   & $2.0\,\times\,10^5$     & 2800 &  & + & 67, 68\\\cutinhead{Radiation diagnostics}
S140                & 22 19 18.2    & +63 18 46.9   & -7.1      & $10^4$                  &                910 & \\
\cutinhead{Disks}
DM Tau   & 04:33:49.7 & $+$18:10:10   & 6.1 & 0.25     & 140  &  &  &  & \tablenotemark{j} \\
LKCa15   & 04:39:17.8 & $+$22:21:04   & 6.0 & 0.74     & 140  &  &  &  &  \\
MWC480   & 04:58:46.3 & $+$29:50:37   & 5.0 & 11.5     & 140  &  &  &  &  \\
TW Hya   & 11:01:51.9 & $-$34:42:17   & 3.0 & 0.25     &  56  &  &  &  &  \\
\tableline
BP Tau   & 04:19:15.8 & $+$29:06:27   & 6.0 & 0.9      & 140  &  &  &  & \tablenotemark{k} \\
GG Tau   & 04:32:30.3 & $+$17:31:41   & 7.0 & 0.8, 0.7 & 140  &  &  &  &  \\
GM Aur   & 04:55:10.2 & $+$30:21:58   & 5.0 & 0.7      & 140  &  &  &  &  \\
MWC758   & 05:30:27.5 & $+$25:19:57   & 5.5 & 21.      & 200  &  &  &  &  \\
T Cha    & 11:57:13.5 & $-$79:21:32   & 5.0 & 0.4      &  66  &  &  &  &  \\
IM Lup   & 15:56:09.2 & $-$37:56:06   & 4.5 & 1.7      & 150  &  &  &  &  \\
AS209    & 16:49:15.3 & $-$14:22:07   & 6.6 & 0.7      & 160  &  &  &  &  \\
HD163296 & 17:56:21.3 & $-$21:57:20   & 6.0 & 35.      & 122  &  &  &  &  \\
\enddata
\tablenotetext{a}{\, Also part of the Outflow sub-program, and will be observed at the given offset coordinates}
\tablenotetext{b}{\, Also part of the Radiation Diagnostics sub-program}
\tablenotetext{c}{\, Low-mass class 0 YSOs}
\tablenotetext{d}{\, Low-mass class I YSOs}
\tablenotetext{e}{\, High-mass prestellar cores}
\tablenotetext{f}{\, mIR-quiet HMPOs}
\tablenotetext{g}{\, mIR-bright HMPOs}
\tablenotetext{h}{\, Hot Molecular Cores}
\tablenotetext{i}{\, UC HII Regions}
\tablenotetext{j}{\, Gas-rich Disks, deep sample}
\tablenotetext{k}{\, Gas-rich Disks, shallow sample}
\tablenotetext{l}{\, Using VLBA observations of a star thought to be associated with the Serpens cluster, \citet{Dzib10} derive a distance of 415 pc}
\tablenotetext{m}{\, Coordinates used in WISH; the more accurate coordinates obtained from SMA observations are 04 39 13.9, $+$25 53 20.6 (TMR1) and 04 39 35.2, $+$25 41 44.4 (TMC1A) \citep{Jorgensen09}}
\tablerefs{(1) \citet{Evans09}; (2) \citet{Jorgensen07}; (3) \citet{Hirota08}; (4) \citet{Jorgensen09}; (5) \citet{Bachiller98}; (6) \citet{Andre00}; (7) \citet{Chen08}; (8) \citet{Seidensticker89}; (9) \citet{Kristensen10}; (10) \citet{vanKempen09h2o}; (11) \citet{Pontoppidan04}; (12) \citet{Hogerheijde99}; (13) \citet{Jorgensen02}; (14) \citet{Hogerheijde97}; (15) \citet{Brinch07}; (16) \citet{Butner91}; (17) \citet{Velusamy07}; (18) \citet{Noriega-crespo04}; (19) \citet{Froebrich05}; (20) \citet{Knude98}; (21) \citet{deGeus90}; (22) \citet{Lommen08}; (23) \citet{Chen09}; (24) \citet{Nisini05}; (25) \citet{Wilking92}; (26) \citet{Gullixson83}; (27) \citet{Mozurkewich86}; (28) \citet{Snell84}; (29) \citet{Stanke00}; (30) \citet{Wilson05}; (31) \citet{Johnstone01}; (32) \citet{Butner90}; (33) \citet{Liseau92}; (34) \citet{Giannini05}; (35) \citet{Slawson88}; (36) \citet{Fuente05pdb}; (37) \citet{Eiroa98}; (38) \citet{Shevchenko89}; (39) \citet{Carey98}; (40) \citet{Pillai06b}; (41) \citet{Beuther02}; (42) \citet{Garay02}; (43) \citet{Beuther08}; (44) \citet{Motte03}; (45) Csengeri et al. in press; (46) \citet{Motte07}; (47) \citet{Rodon08}; (48) \citet{Hachisuka06}; (49) \citet{Ladd93}; (50) \citet{Sridharan02}; (51) \citet{vanderTak00}; (52) \citet{Beuther02b}; (53) \citet{vanderTak06}; (54) \citet{Wyrowski06}; (55) \citet{Bronfman96}; (56) \citet{Sandell00}; (57) \citet{Beuther07}; (58) \citet{Gibba04}; (59) \citet{Beltran05}; (60) \citet{Cesaroni98}; (61) \citet{Mueller02}; (62) \citet{Sollins04}; (63) \citet{Hunter08}; (64) \citet{Olmi96}; (65) Wyrowski (priv. comm.); (66) \citet{Mookerjea07}; (67) \citet{Sandell04}; (68) \citet{deBuizer05}}
\label{tab:sources}
\end{deluxetable}

Recognizing the importance and complexity of the radiative transfer
problem for H$_2$O, the HIFI consortium organized a benchmark workshop
dedicated to an accurate comparison of radiative transfer codes in
2004. The model tests and results are available at a Web
page\footnote{\tt http://www.sron.rug.nl/$\sim$vdtak/H2O} so that
future researchers can test new codes. The reliability of existing
codes has significantly improved thanks to these efforts.  

Predictions of water emission lines have been made for grids of 1D
spherically symmetric envelope models over a large range of
luminosities, envelope masses and other YSO parameters, for a range of
H$_2$O abundances \citep[e.g.,][]{Poelman07,vanKempen08}.  Many of the
line profiles show deep absorptions and horn-like shapes due to the
high line center optical depths discussed above, with emission only
escaping in the line wings.

Chemical models of envelopes and hot cores have been developed
for low- and high-mass sources
\citep[e.g.,][]{Doty02,Doty04,Viti01}. In these models, the physical
conditions are kept static with time. Models in which the physical
structure of the source changes with time as the cloud collapses and
the luminosity evolves have been made by
\citet{Ceccarelli96}, \citet{Viti99}, \citet{Rodgers03}, \citet{Lee04},
\citet{Doty06}, and \citet{Aikawa08} in 1D and
by \citet{Visser09} in 2D.  Grids of 1D $C-$type shock models by
\citet{Kaufman96} and bow-shock models by \citet{Gustafsson10} are
also available.  The effects of high-energy irradiation have been
studied in spherical symmetry in a series of papers 
\citep{Stauber04,Stauber05,Stauber06} and have been extended to
multidimensional geometries
\citep{Bruderer09a,Bruderer09b,Bruderer10mod}.

\subsection{Modeling approach}
\label{sect:model}

Spherically symmetric models of protostellar envelopes such as
illustrated in Fig.~\ref{fig:nt} are constructed for all sources in
the WISH sample by assuming a power-law density profile and
calculating the dust temperature with radius using a continuum
radiative transfer code such as DUSTY \citep{Ivezic97} or RADMC
\citep{Dullemond04model}, taking the central luminosity as input. The
power-law exponent, spatial extent and envelope dust mass are determined
from $\chi^2$ minimization to the far-infrared spectral energy
distribution and submillimeter continuum maps \citep[see procedure
by][]{Jorgensen02}.  The gas temperature is taken to be equal to the
dust temperature, which is a good approximation at these high
densities where gas-dust coupling is significant \citep{Doty97}, and
the gas mass density is obtained through 
multiplication of the dust density by a factor of 100.
These models are termed `passively heated' envelope models, to
distinguish them from `active' shock and UV photon heating mechanisms.

At typical distances of star-forming regions, the warm and cold
regions -- and thus the H$_2$O chemistry zones-- will be largely
spatially unresolved in the {\it Herschel} beams.  However, the abundance
variations (`where is the water spatially along the line of sight?')
can be reconstructed through multi-line, single position observations
coupled with physical models of the sources and radiative transfer
analyses.  This `backward modeling' technique for retrieval of the
abundance profiles in protostellar envelopes has been demonstrated for
ground-based data on a variety of molecules such as CO
\citep{Jorgensen02}, CH$_3$OH
\citep[e.g.,][]{vanderTak00meth,Maret05,Kristensen10meth} and H$_2$CO
\citep[e.g.,][]{vanDishoeck95,Ceccarelli00h2co,Schoier02,Maret05,Jorgensen05h2co}
and confirmed by interferometer data
\citep{Jorgensen04l483,Schoier04}.  Asymmetric rotors like H$_2$O with
many lines from different energies close in frequency are particularly
well suited for such analyses.


\begin{deluxetable}{l r r r r r r}
\tablecolumns{7}
\tabletypesize{\scriptsize}
\tablenum{2}
\tablewidth{0pc}
\tablecaption{WISH Line List}
\tablehead{
\colhead{Species} & \colhead{Transition} & \colhead{$\nu$} & \colhead{$\lambda$} & \colhead{$E_{\rm u}/k_{\rm B}$} & \colhead{$A$} & \colhead{Ref.} \\
\colhead{} & \colhead{} & \colhead{(GHz)} & \colhead{($\mu$m)} & \colhead{(K)} & \colhead{(s$^{-1}$)} \\}
\startdata
\cutinhead{HIFI lines}
H$_2$O      &  1$_{10}$--1$_{01}$  &  556.936  & 538.3  &  61.0  & 3.5($-$3) & 1 \\
            &  2$_{12}$--1$_{01}$  & 1669.905  & 179.5  &  114.4 & 5.6($-$2) & 1 \\
            &  1$_{11}$--0$_{00}$  & 1113.343  & 269.3  &  53.4  & 1.8($-$2) & 1 \\
            &  2$_{02}$--1$_{11}$  &  987.927  & 303.5  & 100.8  & 5.8($-$3) & 1 \\
            &  2$_{11}$--2$_{02}$  &  752.033  & 398.6  & 136.9  & 7.1($-$3) & 1 \\
            &  3$_{12}$--3$_{03}$  & 1097.365  & 273.2  & 249.4  & 1.6($-$2) & 1 \\
            &  3$_{12}$--2$_{21}$  & 1153.127  & 260.0  & 249.4  & 2.6($-$3) & 1 \\
H$_2^{18}$O &  1$_{10}$--1$_{01}$  &  547.676  & 547.4  &  60.5  & 3.3($-$3) & 1 \\
            &  1$_{11}$--0$_{00}$  & 1101.698  & 272.1  &  52.9  & 1.8($-$2) & 1 \\
            &  2$_{02}$--1$_{11}$  &  994.675  & 301.4  & 100.6  & 6.0($-$3) & 1 \\
            &  3$_{12}$--3$_{03}$  & 1095.627  & 273.6  & 248.7  & 1.6($-$2) & 1 \\
H$_2^{17}$O &  1$_{10}$--1$_{01}$  &  552.021  & 543.1  &  60.7  & 3.4($-$3) & 1 \\
            &  1$_{11}$--0$_{00}$  & 1107.167  & 270.8  &  53.1  & 1.8($-$2) & 1 \\
OH\tablenotemark{a}  & $\Omega,J=$1/2,3/2--1/2,1/2  & 1834.747  & 163.4  & 269.8  & 6.4($-$2) & 1 \\
                     & $\Omega,J=$1/2,3/2--1/2,1/2  & 1837.817  & 163.1  & 270.1  & 6.4($-$2) & 1 \\
OH$^+$\tablenotemark{a}      & $N_{J,F}=$1$_{1,3/2}$--0$_{1,3/2}$ & 1033.119  & 290.2  &  49.6  & 1.8($-$2) & 2 \\
            & $N_{J,F}=$2$_{1,3/2}$--1$_{1,3/2}$ & 1892.227  & 158.4  & 140.4  & 5.9($-$2) & 2 \\
H$_2$O$^+$\tablenotemark{a}  & $N_{K_aK_b,J}=$2$_{02,3/2}$--1$_{11,3/2}$ &  746.3  & 401.7  &  89.3  & 5.5($-$4) & 3 \\
            & $N_{K_aK_b,J}=$1$_{11,3/2}$--0$_{00,1/2}$ & 1115.204  & 268.8  &  53.5  & 3.1($-$2) & 6 \\
            & $N_{K_aK_b,J}=$1$_{11,1/2}$--0$_{00,1/2}$ & 1139.6  & 263.1  &  54.7  & 2.9($-$2) & 3 \\
            & $N_{K_aK_b,J}=$3$_{12,5/2}$--3$_{03,5/2}$ &  999.8  & 299.8  & 223.9  & 2.3($-$2) & 3 \\
H$_3$O$^+$  & $J_{K,+}$=0$_{0,-}$--1$_{0,+}$ &  984.712  & 304.4  &  54.7  & 2.3($-$2) & 1 \\
            & $J_{K,+}$=4$_{3,+}$--3$_{3,-}$ & 1031.294  & 290.7  & 232.2  & 5.1($-$3) & 1 \\
            & $J_{K,+}$=4$_{2,+}$--3$_{2,-}$ & 1069.827  & 280.2  & 268.8  & 9.8($-$3) & 1 \\
            & $J_{K,+}$=6$_{2,-}$--6$_{2,+}$ & 1454.563  & 206.1  & 692.6  & 7.1($-$3) & 1 \\
            & $J_{K,+}$=2$_{1,-}$--2$_{1,+}$ & 1632.091  & 183.7  & 143.1  & 1.7($-$2) & 1 \\
CO          &               10--9  & 1151.985  & 260.2  & 304.2  & 1.0($-$4) & 1 \\
            &              16--15  & 1841.346  & 162.8  & 751.7  & 4.1($-$4) & 1 \\
$^{13}$CO   &                 5--4 &  550.926  & 544.2  &  79.3  & 1.1($-$5) & 1 \\
            &                10--9 & 1101.350  & 272.2  & 290.8  & 8.8($-$5) & 1 \\
C$^{18}$O   &                 5--4 &  548.831  & 546.2  &  79.0  & 1.1($-$5) & 1 \\
            &                 9--8 &  987.560  & 303.6  & 237.0  & 6.4($-$5) & 1 \\
            &                10--9 & 1097.163  & 273.2  & 289.7  & 8.8($-$5) & 1 \\
HCO$^+$     &                 6--5 &  535.062  & 560.3  &  89.9  & 1.3($-$2) & 1 \\
CH\tablenotemark{a}          & $J_{F,P}$=3/2$_{2,-}$--1/2$_{1,+}$ &  536.761  & 558.5  &  25.8  & 6.4($-$4) & 2 \\
            & $J_{F,P}$=5/2$_{3,+}$--3/2$_{2,-}$ & 1661.107  & 180.5  & 105.5  & 3.8($-$2) & 2 \\
            & $J_{F,P}$=5/2$_{3,-}$--3/2$_{2,+}$ & 1656.956  & 180.9  & 105.2  & 3.7($-$2) & 2 \\
CH$^+$      &                 1--0 &  835.138  & 359.0  &  40.1  & 6.4($-$2) & 2 \\
            &                 2--1 & 1669.281  & 179.6  & 120.2  & 6.1($-$2) & 2 \\
HCN         &               11--10 &  974.487  & 316.4  & 280.7  & 4.6($-$2) & 1 \\
NH\tablenotemark{a}          & $N_{J,F_1,F}=$1$_{2,5/2,7/2}$--0$_{1,3/2,5/2}$ &  974.478  & 307.6 &  46.8  & 6.9($-$3) & 2 \\
            & $N_{J,F_1,F}=$1$_{1,3/2,5/2}$--0$_{1,3/2,5/2}$ &  999.973  & 299.8  &  48.0  & 5.2($-$3) & 2 \\
NH$^+$\tablenotemark{a}      &    $J_P=$3/2$_-$--1/2$_+$  & 1012.540  & 296.1  &  48.6  & 5.4($-$2) & 7 \\
            &    $J_P=$3/2$_+$--1/2$_-$  & 1019.211  & 294.1  &  48.9  & 5.5($-$2) & 7 \\
NH$_3$      &         1$_0$--0$_0$ &  572.498  & 523.7  &  27.5  & 1.6($-$3) & 1 \\
C$^+$       &  $^2$P$_{3/2}$--$^2$P$_{1/2}$         & 1900.537  & 157.7  &  91.3  & 2.3($-$6) & 5 \\
SH$^+$\tablenotemark{a} & $N_{J,F}=$ 2$_{3,7/2}$--1$_{2,5/2}$ & 1082.909  & 276.8  &  77.2  & 9.8($-$3) & 2 \\
            &  $N_{J,F}=$3$_{4,9/2}$--2$_{3,5/2}$ & 1632.518  & 183.6  & 155.6  & 3.6($-$2) & 2 \\
            &  $N_{J,F}=$1$_{2,5/2}$--0$_{1,3/2}$ &  526.048  & 569.9  &  25.2  & 9.6($-$4) & 2 \\
CS          &               11--10 &  538.689  & 556.5  & 155.1  & 3.3($-$3)  & 1 \\
SH\tablenotemark{a}          &  $N'_\Lambda=$3$_{+1}$--2$_{-1}$  & 1447.012  & 207.2  & 640.6  & 8.1($-$3) & 1 \\
H           &        H(20)$\alpha$ &  764.230  & 401.7  &  \\
\cutinhead{PACS lines}
H$_2$O      & 2$_{21}$--2$_{12}$   & 1661.008  & 180.5  &  194.1 & 3.1($-$2) & 1 \\
            & 2$_{12}$--1$_{01}$   & 1669.905  & 179.5  &  114.4 & 5.6($-$2) & 1 \\
            & 3$_{03}$--2$_{12}$   & 1716.770  & 174.6  &  196.8 & 5.1($-$2) & 1 \\
            & 3$_{13}$--2$_{02}$   & 2164.132  & 138.5  &  204.7 & 1.3($-$1) & 1 \\
            & 3$_{30}$--3$_{21}$   & 2196.346  & 136.5  &  410.7 & 6.6($-$2) & 1 \\
            & 4$_{04}$--3$_{13}$   & 2391.573  & 125.4  &  319.5 & 1.7($-$1) & 1 \\
            & 4$_{14}$--3$_{03}$   & 2640.474  & 113.5  &  323.5 & 2.5($-$1) & 1 \\
            & 2$_{21}$--1$_{10}$   & 2773.977  & 108.1  &  194.1 & 2.6($-$1) & 1 \\
            & 3$_{22}$--2$_{11}$   & 3331.458  &  90.0  &  296.8 & 3.5($-$1) & 1 \\
            & 7$_{07}$--6$_{16}$   & 4166.852  &  71.9  &  843.5 & 1.2(0)\phantom{$-$} & 1 \\
            & 8$_{18}$--7$_{07}$   & 4734.296  &  63.3  & 1070.7 & 1.8(0)\phantom{$-$} & 1 \\
            & 9$_{09}$--8$_{18}$   & 5276.520  &  56.8  & 1323.9 & 2.5(0)\phantom{$-$} & 1 \\
OH     & $\Omega,J=\frac{1}{2}$,$\frac{3}{2}$--$\frac{1}{2}$,$\frac{1}{2}$ & 1834.747  & 163.4  & 269.8  & 6.4($-$2) & 1 \\
       &                                                          & 1837.817  & 163.1  & 270.1  & 6.4($-$2) & 1 \\
       & $\Omega,J=\frac{3}{2}$,$\frac{5}{2}$--$\frac{3}{2}$,$\frac{3}{2}$ & 2509.949  & 119.4  & 120.5  & 1.4($-$1) & 1 \\
       &                                                          & 2514.317  & 119.2  & 120.7  & 1.4($-$1) & 1 \\
       & $\Omega,J=\frac{3}{2}$,$\frac{7}{2}$--$\frac{3}{2}$,$\frac{5}{2}$ & 3543.779  &  84.6  & 290.5  & 5.1($-$1) & 1 \\
       &                                                          & 3551.185  &  84.4  & 291.2  & 5.2($-$1) & 1 \\
       & $\Omega,J=\frac{1}{2}$,$\frac{1}{2}$--$\frac{3}{2}$,$\frac{3}{2}$ & 3786.170  &  79.2  & 181.7  & 3.5($-$2) & 1 \\
       &                                                          & 3789.180  &  79.1  & 181.9  & 3.5($-$2) & 1 \\
CO          &              14--13  & 1611.794  & 186.0  &  580.5 & 2.7($-$4) & 1 \\
            &              16--15  & 1841.346  & 162.8  &  751.7 & 4.1($-$4) & 1 \\
            &              18--17  & 2070.616  & 144.8  &  945.0 & 5.7($-$4) & 1 \\
            &              22--21  & 2528.172  & 118.6  & 1397.4 & 1.0($-$3) & 1 \\
            &              23--22  & 2642.330  & 113.5  & 1524.2 & 1.1($-$3) & 1 \\
            &              24--23  & 2756.388  & 108.8  & 1656.5 & 1.3($-$3) & 1 \\
            &              29--28  & 3325.005  &  90.2  & 2399.8 & 2.1($-$3) & 1 \\
            &              30--29  & 3438.365  &  87.2  & 2564.8 & 2.3($-$3) & 1 \\
            &              31--30  & 3551.592  &  84.4  & 2735.3 & 2.5($-$3) & 1 \\
            &              32--31  & 3664.684  &  81.8  & 2911.2 & 2.7($-$3) & 1 \\
            &              33--32  & 3777.634  &  79.4  & 3092.5 & 3.0($-$3) & 1 \\
            &              36--35  & 4115.606  &  72.8  & 3668.8 & 3.6($-$3) & 1 \\
O           & $^3$P$_1$--$^3$P$_2$         & 4166.852  &  63.2  & 227.7  & 8.9($-$5) & 5 \\
            & $^3$P$_0$--$^3$P$_2$         & 5276.519  & 145.5  & 326.6  & 1.8($-$5) & 5 \\
C$^+$       & $^2$P$_{3/2}$--$^2$P$_{1/2}$ & 1900.537  & 157.7  &  91.3  & 2.3($-$6) & 5 \\
\enddata
\tablenotetext{a}{Fine/hyperfine-structure transition. Only the strongest component is listed here.}
\tablerefs{(1) JPL spectral line catalogue, \citet{Pickett98}; (2) CDMS, \citet{Muller01}; (3) \citet{Bruderer06}; (4) \citet{Bruderer10}; (5) NIST atomic spectra database; (6) \citet{Murtz98}; (7) \citet{Hubers09}}
\label{tab:lines}
\end{deluxetable}

Given a temperature and density structure, the molecular excitation
and radiative transfer in the line is computed at each position in the
envelope.  The resulting sky brightness distribution is convolved with
the beam profile. A trial abundance of water is chosen (see example in
Fig.~\ref{fig:nt}) and is adjusted until the best agreement with
observational data is reached. The velocity structure is represented
either by a turbulent broadening width that is constant with position
or by some function, for example an infall velocity profile. 

The alternative, `forward modeling' approach starts from a full
physico-chemical model and computes the water emission at different
times in the evolution for comparison with observations (see Fig.\ 1
of Doty et al.\ 2004 for a flow-chart of both procedures). In this case,
one obtains best fit model parameters such as the timescale since
evaporation (often labeled as the `age' of the source) or the
cosmic-ray ionization rate \citep[e.g.,][]{Doty06}.

The spherically symmetric models are an important first step but
initial {\it Herschel} results show that they are generally not sufficient
to interpret the data.  Other components such as shocks or UV-heated
cavity walls need to be added in a 2D geometry
\citep{vanKempen10,Visser10}.

\section{Observations}
\label{sect:obs}

\subsection{Source and line selection}
\label{sect:sourceline}

The WISH program contains about 80 sources covering a range of
luminosities and evolutionary stages and uses about 425 hr of {\it
  Herschel} time. The sources are summarized in Table~\ref{tab:sources},
where they are subdivided in a set of subprograms.  Multi-line pointed
observations at the source position are performed using the HIFI and
PACS instruments. In addition, small maps over a few arcmin region are
made for selected sources using either the on-the-fly (HIFI) or raster
mapping (PACS) strategies.  The number of sources per (sub)category
ranges from two for the cold line-poor sources to more than ten for
warm line-rich objects, large enough to allow individual source
pecularities to be distinguished from general trends.  These deep
integrations and thorough coverage of the various types of H$_2$O
lines will set the stage to design future {\it Herschel} programs of
larger, more statistically significant samples using fewer lines.
Specific source selection for each subprogram is discussed in \S 4. More
than 90\% of our sources are visible with ALMA ($\delta <$40$^o$).

The selection of H$_2$O lines observed with HIFI and PACS is
summarized in Fig.~\ref{fig:energy} and Table~\ref{tab:lines} and is
based on information from existing ISO-LWS, SWAS and Odin data, and on
extensive modeling performed for shocks
\citep[e.g.,][]{Neufeld89a,Kaufman96,Giannini06}, low-mass
\citep{Ceccarelli00model,vanKempen08} and high-mass YSO envelopes
\citep{Doty97,Walmsley05,Poelman07}.  Three sets of lines can be
distinguished. The ground-state $o-$ and $p-$H$_2$O lines at 557 and
1113 GHz are the prime diagnostics of the cold gas. These and other
lines connected with ground-state levels (e.g., $2_{12}-1_{01}$ at
1670 GHz) usually show strong self-absorption or they can even be
purely in absorption against the strong continuum provided by the
source itself. The second group are ``medium-$J$'' lines originating
from levels around 100--250 K above ground, which probe the warm gas
(e.g., $2_{02}-1_{11}$ at 988 GHz and $3_{12}-3_{03}$ at 1097 GHz).
The third group are the highly excited lines originating from levels
$>300$~K, which are only populated in strong shocks (e.g.,
higher-lying backbone lines) or which are anomalously excited by
collisional or infrared pumping leading to maser emission.  Because of
high optical depths, optically thin isotopic lines are crucial for the
interpretation so that deep integrations on a number of H$_2^{18}$O
and H$_2^{17}$O lines are included in the program.

For all deeply embedded YSOs, a common strategy is adopted by observing
the ground-state $o-$H$_2$O $1_{10}-1_{01}$ 557 GHz and $p-$H$_2$O
$1_{11}-0_{00}$ 1113 GHz lines, as well as the medium-$J$ $p-$H$_2$O
$2_{02}-1_{11}$ at 988 GHz, $2_{11}-2_{02}$ at 752 GHz, and the
$o-$H$_2$O $3_{12}-2_{21}$ 1097 GHz lines, together with at least one
H$_2^{18}$O, OH and CO line (see Table~\ref{tab:lines}).
Several lines of $^{13}$CO, C$^{18}$O and hydrides are covered
serendipitously in the same settings.  Higher-$J$ H$_2$O, OH, [O I]
and CO lines are probed with a series of PACS settings on all low- and
intermediate-mass sources, which provide $5\times 5$ mini-maps on a
$47''\times 47''$ scale with 9.4$''$ spacing to probe extended emission. 
The PACS data also give information on the dust continuum emission at
wavelengths poorly sampled to date. For high-mass YSOs and a few
bright low-mass YSOs, full PACS spectral scans are taken, which cover
many high-excitation H$_2$O lines in an unbiased way.  Integration
times are generally such that H$_2$O abundances down to $\sim
10^{-9}-10^{-10}$ are probed, so that regions with significant
freeze-out can be studied. Typically 5--6 hr are spent per source to
cover all lines with integration times per line ranging from 10
minutes to 1 hr. For the pre-stellar cores and protoplanetary disks,
long integration times with HIFI of up to 12 hr per line are chosen to
ensure that any non-detections are significant. These are the deepest
observations carried out with HIFI across all KPs.

\subsection{Observations and data reduction}

\subsubsection{HIFI}

The bulk of the HIFI data are taken in dual-beam switch mode with a
nod of $3'$ using fast chopping.  The HIFI receivers are double
sideband with a sideband ratio close to unity. For line-rich sources
(mostly high-mass YSOs), the local oscillator was shifted slightly for
half of the integration time to disentangle lines from the upper and
lower side bands. Two polarizations, H and V, are measured
simultaneously and are generally averaged together to improve
the signal-to-noise. In some cases, differences of order 30\% are found
between the two polarizations in which case only the higher quality
H-band spectra are used for analysis since the mixers have been
optimized for H band.  Two back-ends are employed: the low-resolution
back-end (WBS) with an instantaneous bandwidth of 4 GHz at 1.1 MHz
spectral resolution ($\sim$1100 and 0.3 km s$^{-1}$ at 1 THz, respectively) 
and the
high-resolution back-end (HRS) with a variable bandwidth and
resolution (typically 230 MHz bandwidth and 250 kHz resolution, or
$\sim$63 and 0.07 km s$^{-1}$ at 1 THz, respectively).

Data reduction of the HIFI spectra involves the usual steps of
checking individual exposures for bad spectra, summing exposures,
taking out any baseline ripples, fitting a low-order polynomial
baseline, and making Gaussian fits or integrating line intensities as
appropriate.  The data are reduced within the Herschel Interactive
Processing Environment (HIPE) (Ott 2010) and can be exported to
CLASS\footnote{\tt http://www.iram.fr/IRAM/GILDAS} after level 2 for
further analysis. The main beam efficiency has been determined to be
around 0.76 virtually independent of frequency, except for a 15\%
lower value around 1.1--1.2 THz \citep{Olberg10}, and the absolute
calibration is currently estimated to be better than $\sim$15\% for
HIFI Bands 1, 2 and 5, and $\sim$30\% for Bands 3, 4, 6 and 7.
The rms noise in the WBS data is generally lower than that in
the HRS data by a factor of 1.4 when binned to the same spectral
resolution, due to a $\sqrt 2$ loss factor in the HRS autocorrelator.

\subsubsection{PACS}

PACS is a 5$\times$5 array of 9.4$''$$\times$9.4$''$ spaxels (spatial
pixels) with very small gaps between the pixels. Each spaxel covers
the 53--210 $\mu$m wavelength range with a spectral resolving power
ranging from 1000 to 4000 (the latter only below 63 $\mu$m) in
spectroscopy mode. In one exposure, a wavelength segment is observed
in the first order (105--210 $\mu$m) and at the same time in the
second (72--105 $\mu$m) or third order (53--72 $\mu$m).  Two different
nod positions, located 6$'$ in opposite directions from the target,
are used to correct for telescopic background.  Data are reduced
within HIPE.  The uncertainty in absolute and relative fluxes is
estimated to be about 10--20\%, based on comparison with the ISO-LWS
data.

The diffraction-limited beam is smaller than a spaxel of 9.4$''$ at
wavelengths shortward of 110 $\mu$m. At longer
wavelengths, the point-spread function (PSF) becomes significantly
larger such that at 200 $\mu$m only 40\% of the light of a well
centered point source falls on the central pixel.  Even at 100 $\mu$m,
30\% of the light still falls outside the central spaxel. Thus,
observed fluxes reported for a single pixel have to be corrected for
the point-source PSF using values provided by the Herschel Science
Center.  For extended sources, fluxes can be summed over the spaxels
to obtain the total flux.  In case of a bright central source with
extended emission, such as along an outflow, fluxes at the outflow
positions were corrected for the leaking of light from the central
spaxel into adjacent spaxels.

\subsection{Archival value}

All reduced HIFI and PACS spectra will be delivered to the Herschel
Science Center approximately one year after the last KP data have been
taken, assuming that reliable calibration is available in a timely
manner.  The delivery will consist of reduced and calibrated HIFI
spectra, line profile parameters (either Gaussian fits or integrated
intensities, depending on line) and reduced small maps, most likely in
FITS format.  A quick-look spectral browser will be on the team
website.  For PACS, a data cube of calibrated images will be
delivered, similar to that for integral field spectrometers.

Over the past decade, the WISH team has collected a wealth of
complementary data on various molecular lines for the majority of our
sources using the JCMT, IRAM 30m, APEX, CSO, Onsala and other
single-dish telescopes.  These data include tracers of the cloud
velocity structure and column density ($^{12}$CO, $^{13}$CO, C$^{18}$O
$J$=1--0, 2--1, and 3--2, probing $\sim 30$~K gas), high temperature
lines (CO and isotopologues 6--5, 7--6, probing $\sim$50--100 K gas)
and high density tracers (CS and C$^{34}$S 2--1, 5--4, 7--6; HCO$^+$
and H$^{13}$CO$^+$ 3--2, 4--3, probing $n>10^5$ cm$^{-3}$), as well as
the atomic [C I] line (probing UV radiation).  Moreover, grain surface
chemistry products are observed through various H$_2$CO and CH$_3$OH
lines, whereas shocks powerful enough to disrupt grain cores are
traced through SiO emission. Examples of such studies are
\citet{vanderTak00,Jorgensen04inv,Fuente05,Nisini07,Marseille08}.

A recent development is the use of heterodyne array receivers such as
the 9 pixel HERA 230 GHz array at the IRAM 30m; the HARP-B 16 pixel
345 GHz receiver at the JCMT
\citep[e.g.,][]{vanKempen09oph,Kristensen10meth}; and the
dual-frequency 2$\times$7 pixel CHAMP+ 650/850 GHz receiver at APEX
\citep[e.g.,][]{vanKempen09hh46,vanKempen09champ2}. The spatial
resolution of these telescopes is typically 7--15$''$, i.e.,
comparable to that of {\it Herschel} at the higher frequencies. These
data also provide an `image' of the warm dense gas contained within
the $\sim 37''$ {\it Herschel} beam at 557 GHz.  Millimeter
interferometry data down to $\sim$1$''$ resolution also exist for a
significant subset of the sources in various molecules and lines
\citep[e.g.,][and references
  cited]{Fuente05pdb,vanderTak06,Benz07,Jorgensen09}. Except for the
compact hot core, however, such data cannot be compared directly with
single dish data from {\it Herschel} since a significant fraction of
the extended emission is resolved out in the interferometer.

Finally, submillimeter continuum maps obtained with JCMT-SCUBA,
IRAM-MAMBO, and/or APEX-LABOCA are available at $10-20''$ resolution
for most embedded sources \citep[e.g.,][]{diFrancesco08}, as are
ground-based and mid-IR {\it Spitzer} observations of ices
\citep{Gibb04,Boogert08}.  {\it Spitzer-IRS} mid-infrared H$_2$ maps
have been taken for the three outflow targets studied in detail
\citep{Neufeld09}.

The ground-based complementary data will be summarized in a table at
the KP Web site.  For several subprograms, the most relevant data will
be available directly at the KP web site.  Specifically, in
collaboration with the `Dust, Ice and Gas in Time' (DIGIT) KP, a
searchable web-based archive of complementary submillimeter
single-dish, interferometer and infrared spectra is being built for
low-mass YSOs.

\begin{figure}
\includegraphics[angle=0,width=0.6\textwidth]{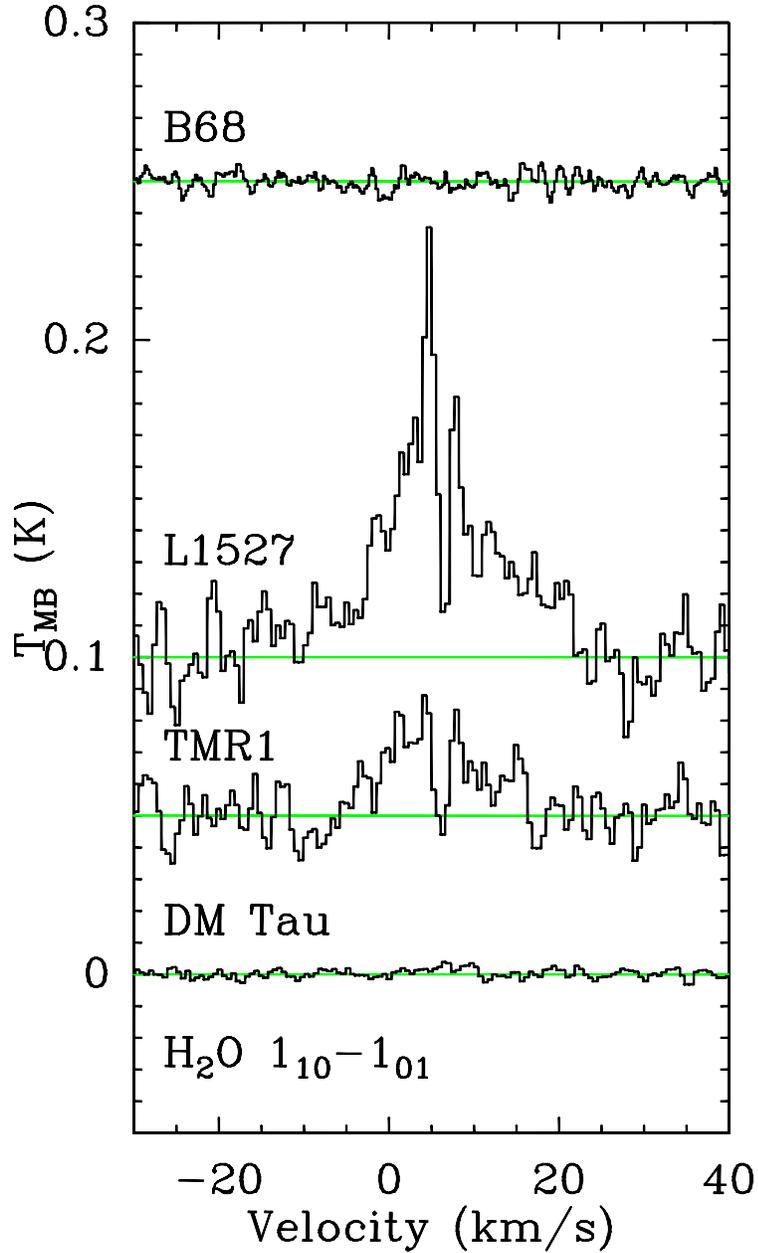}
\caption{HIFI spectra of the $o-$H$_2$O $1_{10}-1_{01}$ 557 GHz line.
  From top to bottom: the pre-stellar core B68
  \citep{Caselli10}, the low-mass Stage 0 YSO L1527, the low-mass
  Stage I YSO TMR-1, and the protoplanetary disk DM Tau
  \citep{Bergin10}.}\label{fig:557spectra}
\end{figure}

\section{Results}

In the following, a brief description of the scientific motivation and
observing strategy for each of the subprograms is given together with
initial results. Full source and line lists are given in
Tables~\ref{tab:sources} and \ref{tab:lines}.  Unless otherwise
specified the cited noise limits are for 0.2--0.4 km s$^{-1}$ velocity
bins and all stated results refer to main beam antenna temperatures,
$T_{MB}$.

\smallskip
\noindent
\subsection{Pre-stellar cores}
\label{sect:prestellar}

{\it Motivation.} In recent years, a number of cold, highly extincted
clouds have been identified which have a clear central density
condensation \citep[see overview by][]{Bergin07}.  Several of these
so-called pre-stellar cores are thought to be on the verge of collapse
and thus represent the earliest stage in the star-formation process
\citep[e.g.,][]{Tafalla98, Caselli02,Crapsi05}. It is now widely
accepted that most molecules are highly depleted in the inner dense
parts of these cores \citep[e.g.,][]
{Caselli99,Bergin02,Tafalla06}. H$_2$O should be no exception,
although its formation as an ice already starts in the general 
molecular cloud phase. The physical properties of these cores are very
well constrained from ground-based continuum and line data and they
are the most pristine types of clouds, without any internal heat
sources and almost no turbulence.  Thus, they form the best
laboratories to test chemical models. Since H$_2$O is the dominant ice
component and has a very different binding energy and formation route
than CO, HIFI observations of H$_2$O will form a critical benchmark to
test gas-grain processes.

HIFI will be key to address the questions: (i) Where does the onset
of H$_2$O ice formation and freeze-out occur in dense pre-stellar
cores?  (ii) How effective are non-thermal desorption mechanisms
(photodesorption at edge, cosmic-ray induced desorption in center) in
maintaining some fraction of O and H$_2$O in the gas phase? (iii) Is
the H$_2$O abundance in the gas phase sensitive to the pre-stellar
core environment?

{\it Source and line selection:} Two well-studied pre-stellar cores,
B68 and L1544, are observed in the $o-$H$_2$O 557 GHz line at the
central position down to $\sim$2 mK rms. Both are at $d<$200 pc and
have bright millimeter continuum dust emission, a simple morphology
and centrally concentrated density profiles.  B68 is an isolated dark
globule exposed to the general interstellar radiation field whereas
L1544 is a dense core embedded in a larger cloud.  The sensitivity
limits are based on H$_2$O column densities estimated using the model
results of \citet{Aikawa05} for chemically `old' sources like
L1544. From the predicted abundance profile and the density structure,
column densities have been determined at different impact parameters
and convolved with the HIFI beam at 557 GHz. The brightness
temperatures have been derived using the non-LTE radiative transfer
codes of \citet{Hogerheijde00} and \citet{Keto04}. The results are
very sensitive to the temperature structure of the core, the initial
H$_2$O abundance in the chemical models, as well as the assumed
  ortho/para ratio of both H$_2$O and H$_2$. Pre-{\it Herschel} model
predictions ranged from peak antenna temperatures of below 10 mK
  \citep{Hollenbach09} to several hundred mK.

\begin{figure}
\includegraphics[angle=0,width=0.5\textwidth]{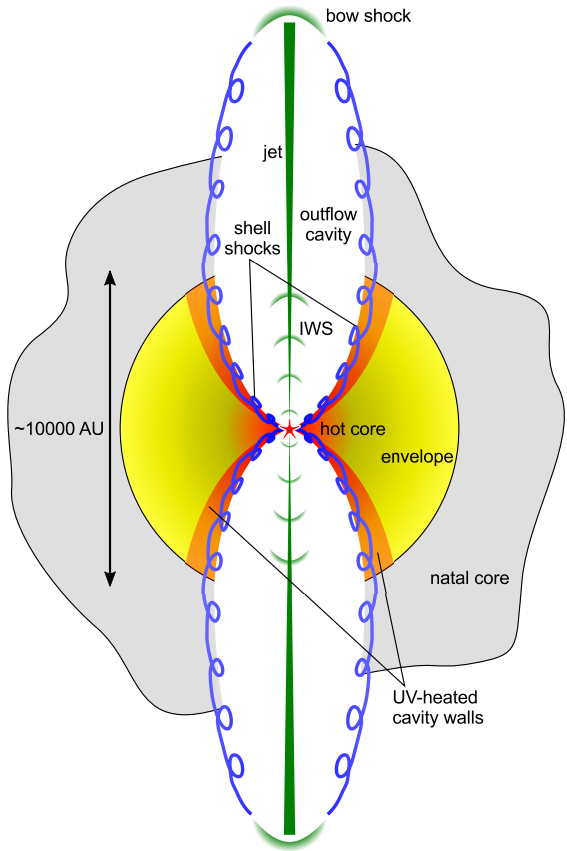}
\caption{Cartoon of a protostellar envelope with the different
  physical components and their nomenclature indicated. IWS stands for
  internal working surfaces. The indicated scale is appropriate for a
  low-mass YSO \citep{Visser10}. On this scale, the $\sim$100 AU radius
  disk surrounding the protostar is not visible.}\label{fig:cartoon}
\end{figure}

{\it Initial results.} \citet{Caselli10} present early results for the
two cores, with the B68 $o$-H$_2$O $1_{10}-1_{01}$ spectrum included
in Fig.~\ref{fig:557spectra}. These spectra took $\sim$4 hr
integration (on+off) each. No emission line is detected in B68 down to
2.0 mK rms in 0.6 km s$^{-1}$ bins. This is more than an order of
magnitude lower than the previous upper limit obtained by SWAS for
this cloud \citep{Bergin02h2o} as well as most pre-{\it Herschel}
model predictions. The 3$\sigma$ upper limit corresponds to an
$o-$H$_2$O column density $<2.5 \times 10^{13}$ cm$^{-2}$ and a mean
line-of-sight abundance $< 1.3 \times 10^{-9}$.  This is
  consistent with the non detection predicted by the model of
\citet{Hollenbach09} applied to B68.

Interestingly, the L1544 spectrum shows a tentative {\it absorption}
feature against the submillimeter continuum, which closely matches the
velocity range spanned by the CO 1--0 emission line. If the water
absorption is confirmed by deeper integrations which are currently
planned within WISH, it provides a very powerful tool to determine the
water abundance profile along the line of sight.

Initial radiative transfer analyses, based on the \citet{Keto10}
models, imply an H$_2$O abundance profile for both clouds which peaks
at an abundance of $\sim 10^{-8}$ in the outer region of the core,
where the density is too low for significant freeze-out but where the
gas is sufficiently shielded from photodissociation ($A_V>2$ mag).
The abundance drops by more than an order of magnitude in the central
part due to freeze-out.  These new predictions are based on an improved
physical structure of the L1544 core and on an H$_2$O abundance
profile which includes the freeze-out, a generalized desorption from
\citet{Roberts07} and photodissociation, but no detailed gas phase and
surface chemistry. Such abundance profiles differ from those expected
from chemical models by \citet{Aikawa05}, who predict higher H$_2$O
abundances toward the core center and thus brighter lines.
Investigations are currently under way to understand which processes
are producing the different abundance structures.

These results demonstrate that questions (i) and (ii) can
be addressed by the {\it Herschel} observations, provided the
integrations are deep enough. Question (iii) requires observations of
a larger sample of cores besides those studied in WISH.

\subsection{Low-mass young stellar objects}
\label{sect:lowmass}

\begin{figure}
\begin{minipage}{0.475\textwidth}
\includegraphics[angle=0,width=\textwidth]{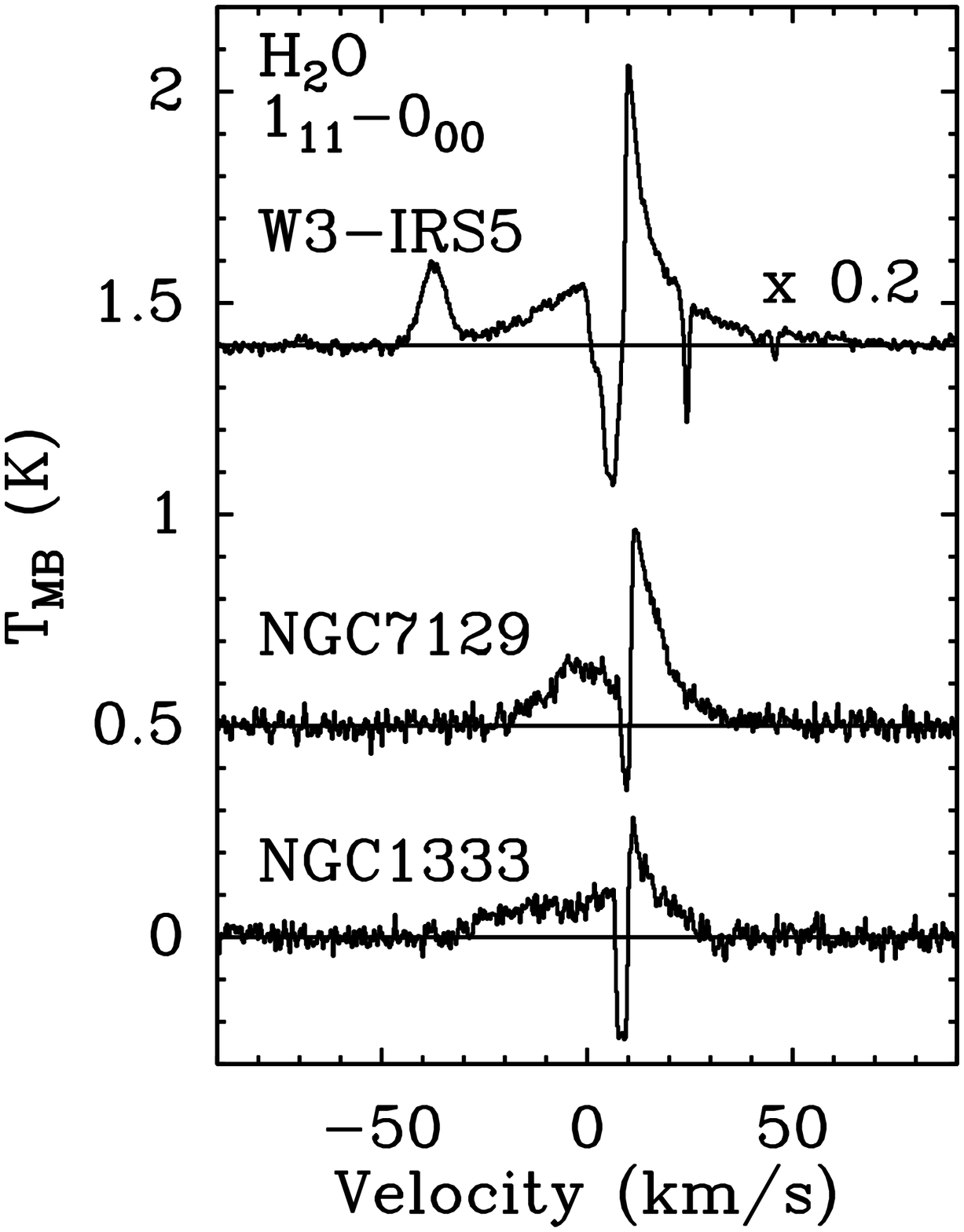}
\end{minipage}\hfill%
\begin{minipage}{0.475\textwidth}
\includegraphics[angle=0,width=\textwidth]{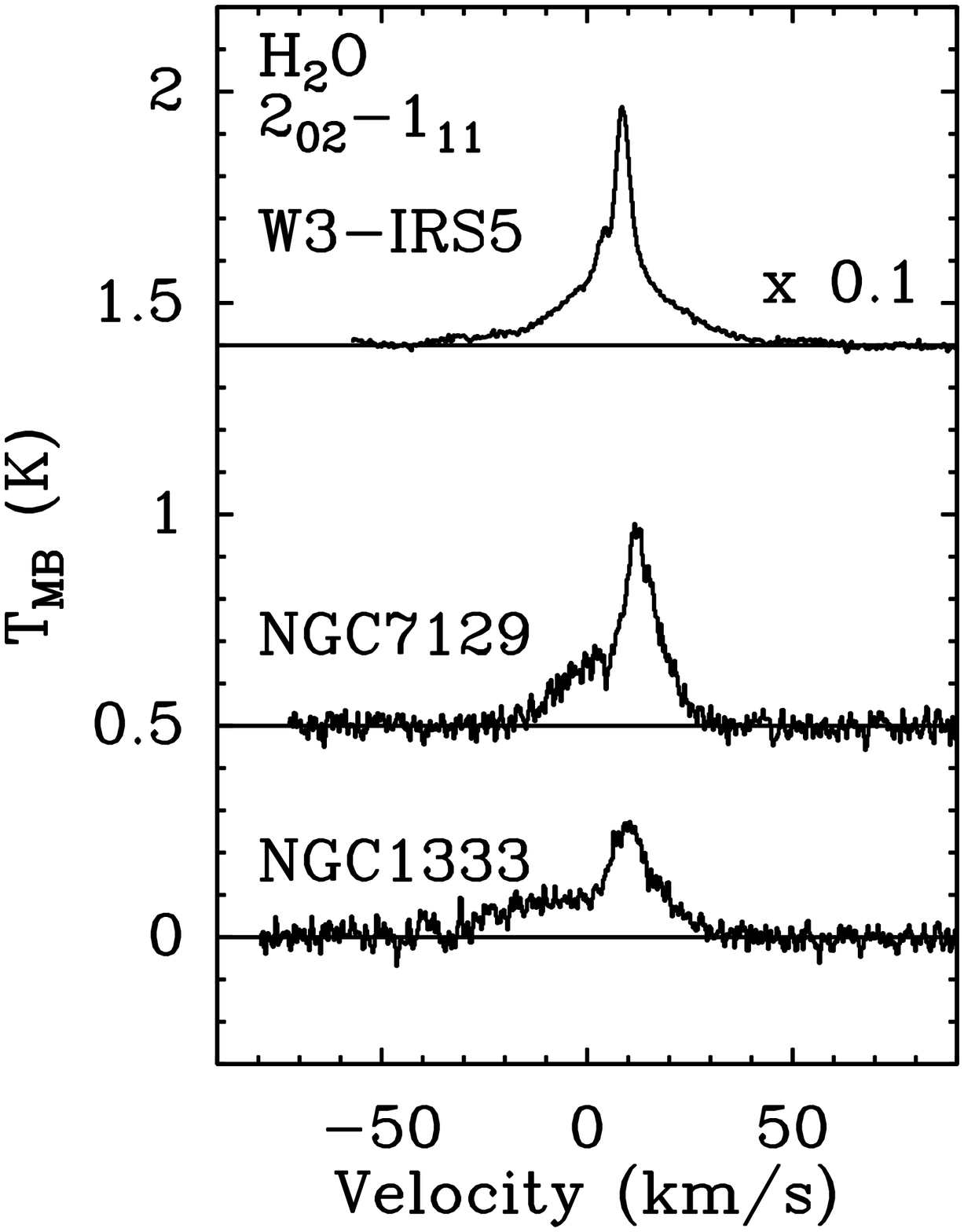}
\end{minipage}\hfill%
\caption{HIFI spectra of the $p-$H$_2$O $1_{11}-0_{00}$ 1113 GHz
  (left) and $2_{02}-1_{11}$ 988 GHz (right) lines.  From top to
  bottom: the high-mass YSO W3 IRS5 ($L=1.5\times 10^5$ L$_\odot$,
  $d=$2.0 kpc) \citep{Chavarria10}, the intermediate-mass YSO NGC 7129
  FIRS2 ($430$ L$_\odot$, 1260 pc) \citep{Johnstone10}, and the
  low-mass YSO NGC 1333 IRAS2A ($20$ L$_\odot$, 235 pc)
  \citep{Kristensen10}.  All spectra have been shifted to a central
  velocity of 0 km s$^{-1}$. The red-shifted absorption features 
  seen in the $1_{11}-0_{00}$ spectrum toward W3 IRS5
  are due to water in foreground clouds whereas the small emission
  line on the blue side  can be ascribed to the SO$_2$
  $13_{9,5}-12_{8,4}$ transition.}
\label{fig:paraspectra}
\end{figure}

{\it Motivation.} Once the cloud has started to collapse and a
protostellar object has formed, the envelope is heated from the inside
by the accretion luminosity. The physical structure of these 
low-mass deeply embedded objects ($\lapprox$ 100 L$_{\odot}$) is complex,
with large gradients in temperature and density in the envelope, a
compact disk at the center, and stellar jets and winds creating a
cavity in the envelope and affecting the surrounding material
\citep[e.g.,][]{vanDishoeck95,Jorgensen05i2,Chandler05} (see
Fig.~\ref{fig:cartoon}).
The high spatial and spectral resolution at
far-infrared wavelengths offered by {\it Herschel} 
are key to disentangling these processes.

Several evolutionary stages can be distinguished for low-mass YSOs
\citep[see nomenclature by][]{Robitaille06}.  In the earliest deeply
embedded Stage 0, the main accretion phase, the envelope mass is still
much larger than that of the protostar or disk ($M_{\rm env}>>M_*$ and
$>>M_{\rm disk}$).  At the subsequent Stage 1, the star is largely
assembled and now has a much larger mass than the disk ($M_*>>M_{\rm
disk}$), but the envelope still dominates the circumstellar material
($M_{\rm env}>M_{\rm disk}$). At Stage 2, the envelope has completely
dissipated and only a gas-rich disk is left around the young pre-main
sequence star. Observationally, low-mass YSOs are usually classified
based on their SEDs according to their spectral slopes (as in the Lada
classes) or bolometric temperatures \citep[see summary in][]{Evans09}.
The Class 0 and II sources generally correspond well to the
evolutionary Stage 0 and 2 cases. In contrast, the traditional Class I
sources are found to be a mix of true Stage 1 sources and of Stage 2
disks which are either seen edge-on or which have a large amount of
foreground extinction resulting in a rising infrared spectrum
\citep[see discussion in][]{Crapsi08}.

ISO-LWS observations have
shown that low-mass YSOs are copious
water emitters, with more than 70\% of the Stage 0 sources
detected in H$_2$O lines with energies up to 500 K above ground
\citep[see summary by][]{Nisini02}.  
Two scenarios have been put forward to explain these strong lines.  In
one model, the lines arise primarily from the bulk of the envelope,
with the lower-energy lines originating in the outer envelope and the
higher-energy lines from the inner hot core \citep[][]{Maret02}. In the
alternative interpretation, most of the H$_2$O emission is associated
with the outflows impacting the surrounding gas on larger scales
\cite[e.g.,][]{Giannini01}.

In contrast, none of the low-mass Stage 1
sources show H$_2$O lines in the ISO-LWS beam \citep{Giannini01}.
Possible explanations given in that paper include dissociation
resulting from enhanced penetration of X-rays or UV photons or a
longer timescale enhancing freeze-out.  HIFI can distinguish between
the different scenarios by providing much deeper searches than
ISO-LWS, as well as the photodissociation products OH and O, and X-ray
products such as ions. Indeed, OH/H$_2$O abundance ratios are
expected to change by factors of $>$100 due to UV and X-rays
\citep{Stauber06}. Complementary infrared data on H$_2$O ice are
available for many of these sources so that the gas/ice ratio can be
measured directly.

Key questions to be addressed are: (i) What is the origin of the strong
H$_2$O emission in low-mass Stage 0 protostars? Passively heated
envelope or shocked outflow material?  What is the role of the outflow
cavity? (ii) Is the high water abundance governed by high-temperature
chemistry or grain mantle evaporation?  (iii) Do low-mass Stage 1
sources indeed have much lower H$_2$O abundances? If so, is this due
to enhanced enhanced UV, X-rays or other effects such as lower outflow
activity?  (iv) What is the HDO/H$_2$O ratio in the protostellar
phase, and what does this tell us about the efficiency of grain
surface deuteration?  How does HDO/H$_2$O evolve during star
formation? Is it similar to that found in comets?

{\it Source and line selection.}  A set of 14 deeply embedded Stage 0
sources has been selected from the list of \citet{Andre00} and 15
Stage 1 sources selected from the lists of \citet{Tamura91}, \citet{Andre94}
and the spectroscopic part of the Spitzer `Cores to Disks' legacy
program \citep{Evans03,Evans09}.  All sources have $d<450$ pc and are
well characterized with a variety of other single-dish submillimeter
telescopes and interferometers \citep{Jorgensen04inv,Jorgensen07}.  In
particular, only bonafide Stage 1 embedded YSOs are included, with the
stage verified through complementary single-dish and interferometer
data \citep{vanKempen09oph,Jorgensen09}.

The line selection for the Stage 0 sources follows the general
philosophy outlined in \S~\ref{sect:sourceline}, with typical
integration times for the HIFI lines such that 20--100 mK rms is
reached in $\sim$0.5 km s$^{-1}$ bins depending on the line, and down
to 5 mK for H$_2^{18}$O. For reference, observed widths of molecular
lines originating in the bulk of the envelopes are typically
1--2 km s$^{-1}$ FWHM. The noise limits are based on 1D envelope
models by \citet{vanKempen08} for L483, one of the weaker Stage 0
sources, and do not include any outflow contribution.  Not all lines
are observed for all sources: some time-consuming higher frequency
lines are dropped for the weaker ones.  Selected PACS lines are
observed to cover the higher excitation emission and obtain
information on the spatial extent.  Full PACS spectral scans are made
of four Stage 0 sources (NGC 1333 IRAS 4A/B, IRAS2A, and Serpens SMM1)
to obtain an unbiased view of the higher H$_2$O lines.
For the Stage 1 objects, only four H$_2$O lines and one CO
line are observed with HIFI. The set of PACS lines observed for these
sources is also more limited.

\begin{figure}
\begin{minipage}{0.5\textwidth}
\includegraphics[angle=0,width=\textwidth]{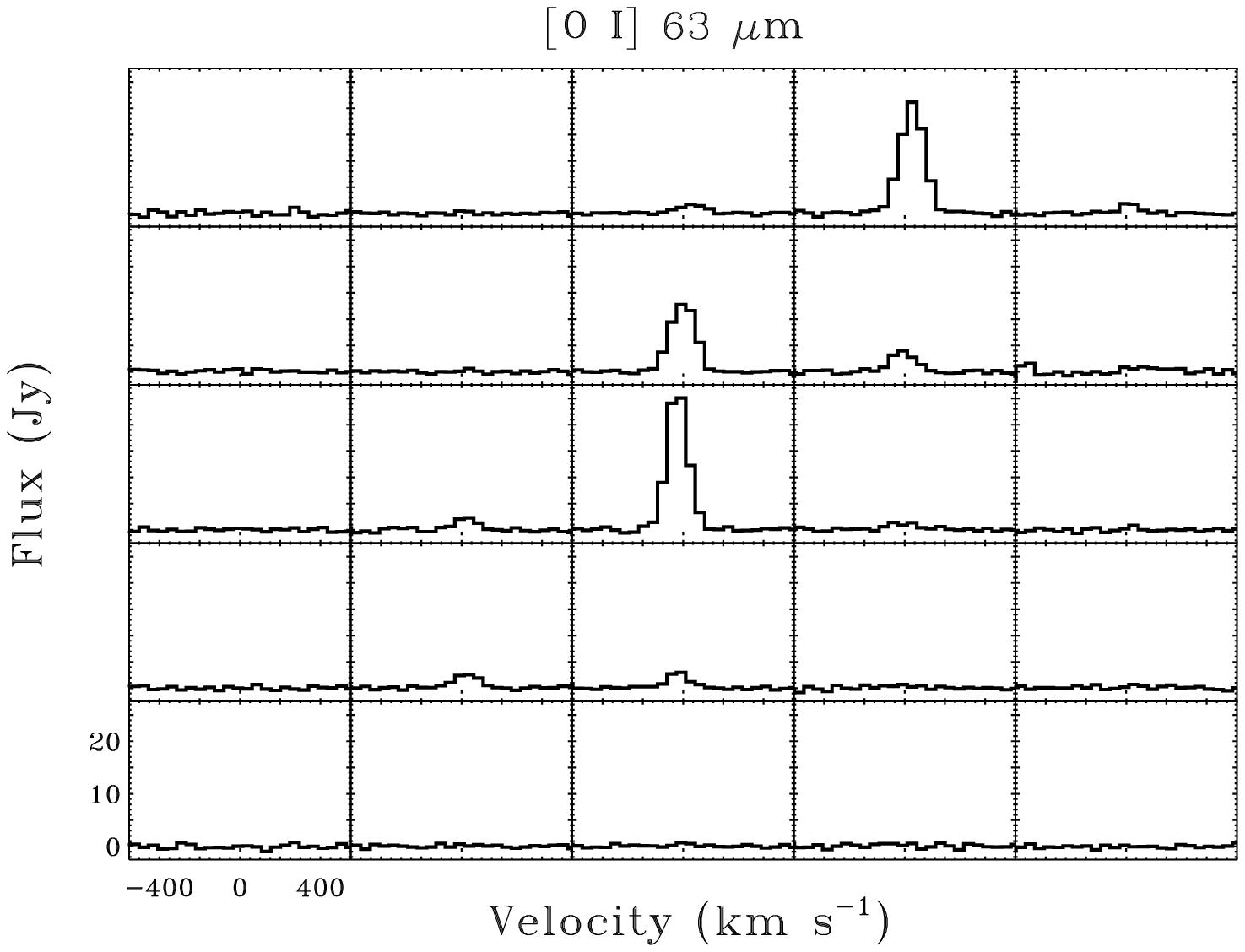}
\end{minipage}\hfill%
\begin{minipage}{0.5\textwidth}
\includegraphics[angle=0,width=\textwidth]{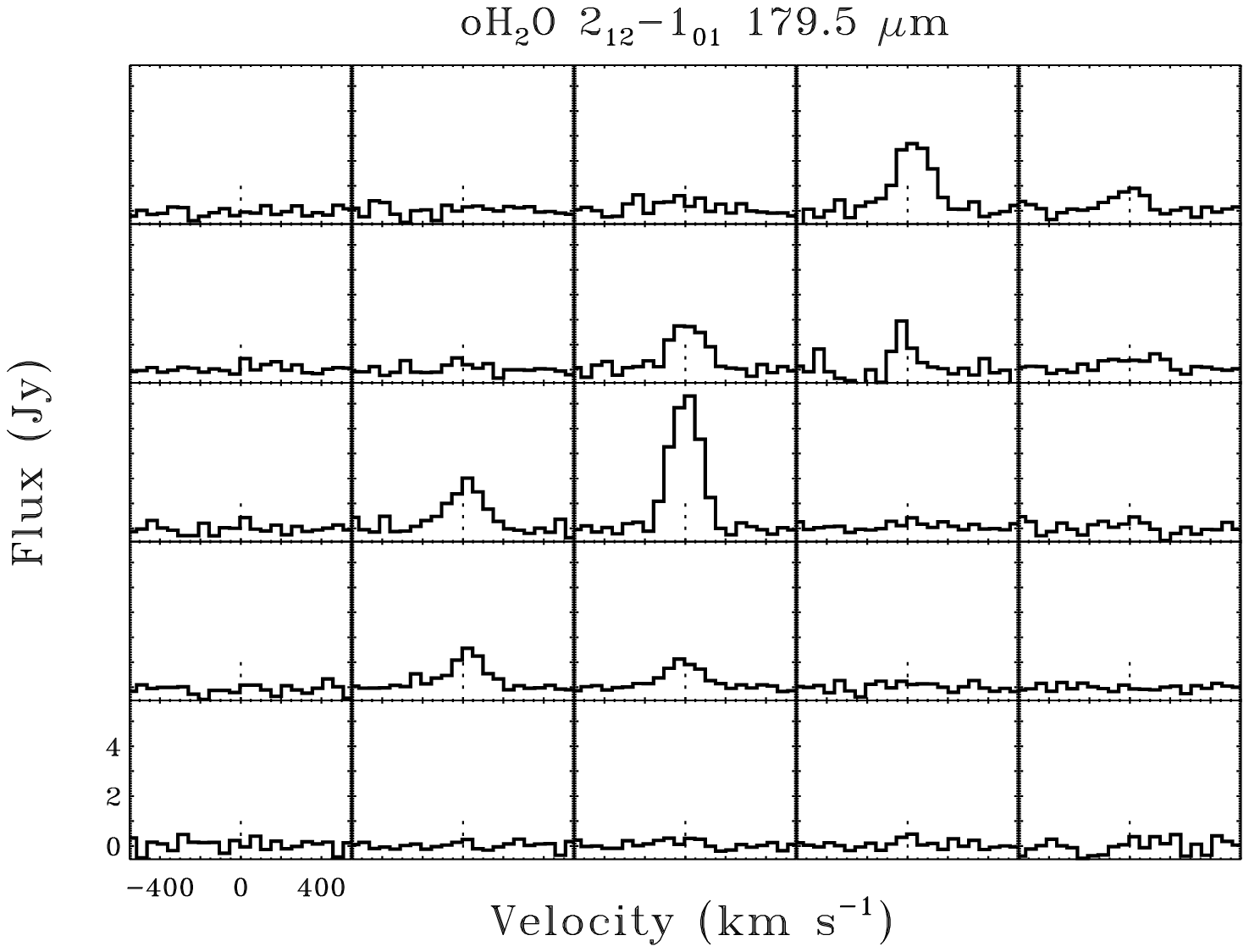}
\end{minipage}\hfill%
\begin{minipage}{0.5\textwidth}
\includegraphics[angle=0,width=\textwidth]{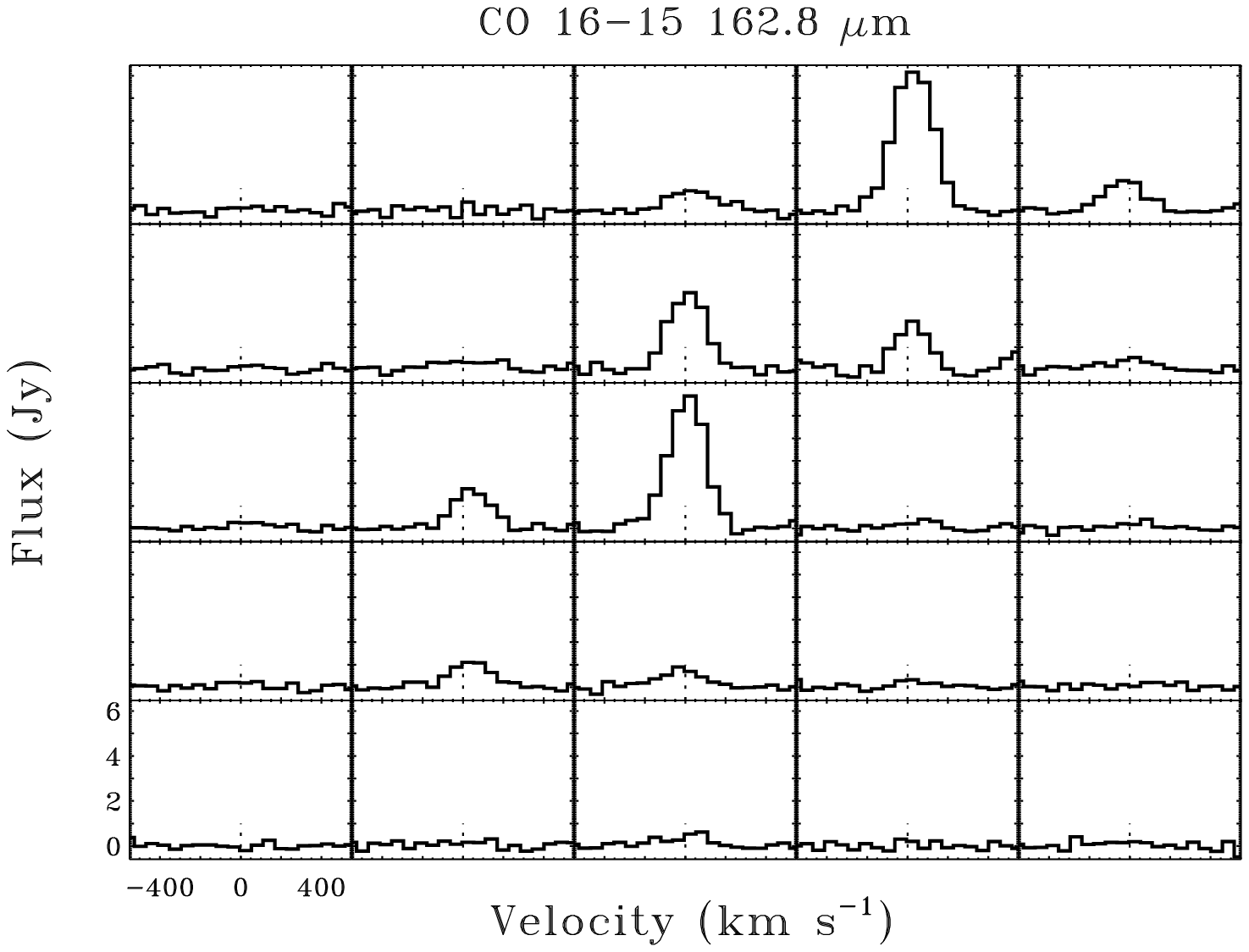}
\end{minipage}\hfill%
\caption{PACS 5$\times$5 spectral maps of the [O I] $^3P_1-^3P_2$ 63
  $\mu$m (left), H$_2$O $2_{12}-1_{01}$ 179 $\mu$m (middle) and CO
  $J$=16--15 163 $\mu$m lines (right) toward the low-mass Stage 1 YSO
  IRAS 15398-3359. All lines are extended along the outflow
  direction seen in the low-$J$ CO map of \citet{vanKempen09south}. }
\label{fig:classIpacs}
\end{figure}

 {\it Initial results.}  Figure~\ref{fig:paraspectra} presents spectra
 of the lowest two $p$-H$_2$O lines toward one of the brightest
 low-mass Stage 0 sources, NGC 1333 IRAS2A \citep{Kristensen10},
 whereas Fig.~\ref{fig:557spectra} includes the lowest $o$-H$_2$O line
 toward the Stage 0 source L1527. The profiles are surprisingly broad
 and complex. They can be decomposed into three components: a broad
 ($>$20 km s$^{-1}$), medium-broad ($\sim$5-10 km s$^{-1}$) and narrow
 (2--3 km s$^{-1}$) component, with the latter seen primarily in
 (self-)absorption in the ground-state transitions.  The central
 velocities of the three components can vary relative to each other
 from source-to-source, probably due to different viewing angles and
 geometries of the systems. For one Stage 0 source, NGC 1333 IRAS4A,
 the $2_{02}-1_{11}$ line shows an inverse P-Cygni profile indicative
 of large scale infall.  Interestingly, the H$_2^{18}$O lines only
 reveal the broad component.

 From the profiles themselves, it can be concluded immediately that
 the bulk of the water emission arises from shocked gas and not in the
 quiescent envelope: a hot core region with an enhanced H$_2$O
   abundance and an infall velocity profile gives lines with widths of
   at most 4--5 km s$^{-1}$ \citep{Visser10}. The broad component is
 associated with shocks on large scales along the outflow cavity
 walls, as traditionally probed with the standard high velocity CO
 outflows (labelled as `shell shocks' in Fig.~\ref{fig:cartoon}). The
 medium component is linked to shocks on smaller scales in the dense
 inner envelope ($<1000$ AU), based on comparison of the line profiles
 with those of high density tracers and with interferometer data
 \citep{Kristensen10}. The H$_2$O/CO abundance ratios in the shocks
 are high, $\sim$0.1--1, corresponding to H$_2$O abundances of $\sim
 10^{-5}-10^{-4}$ with respect to H$_2$. The highest abundances are
 found at the highest velocities, indicating that only a few \% of the
 gas is processed at high enough temperatures where all O is driven
 into H$_2$O.  This small percentage is consistent with the findings
 of \citet{Franklin08} based on SWAS data.

\begin{figure}[]
\begin{minipage}{0.5\textwidth}
\includegraphics[angle=0,width=\textwidth]{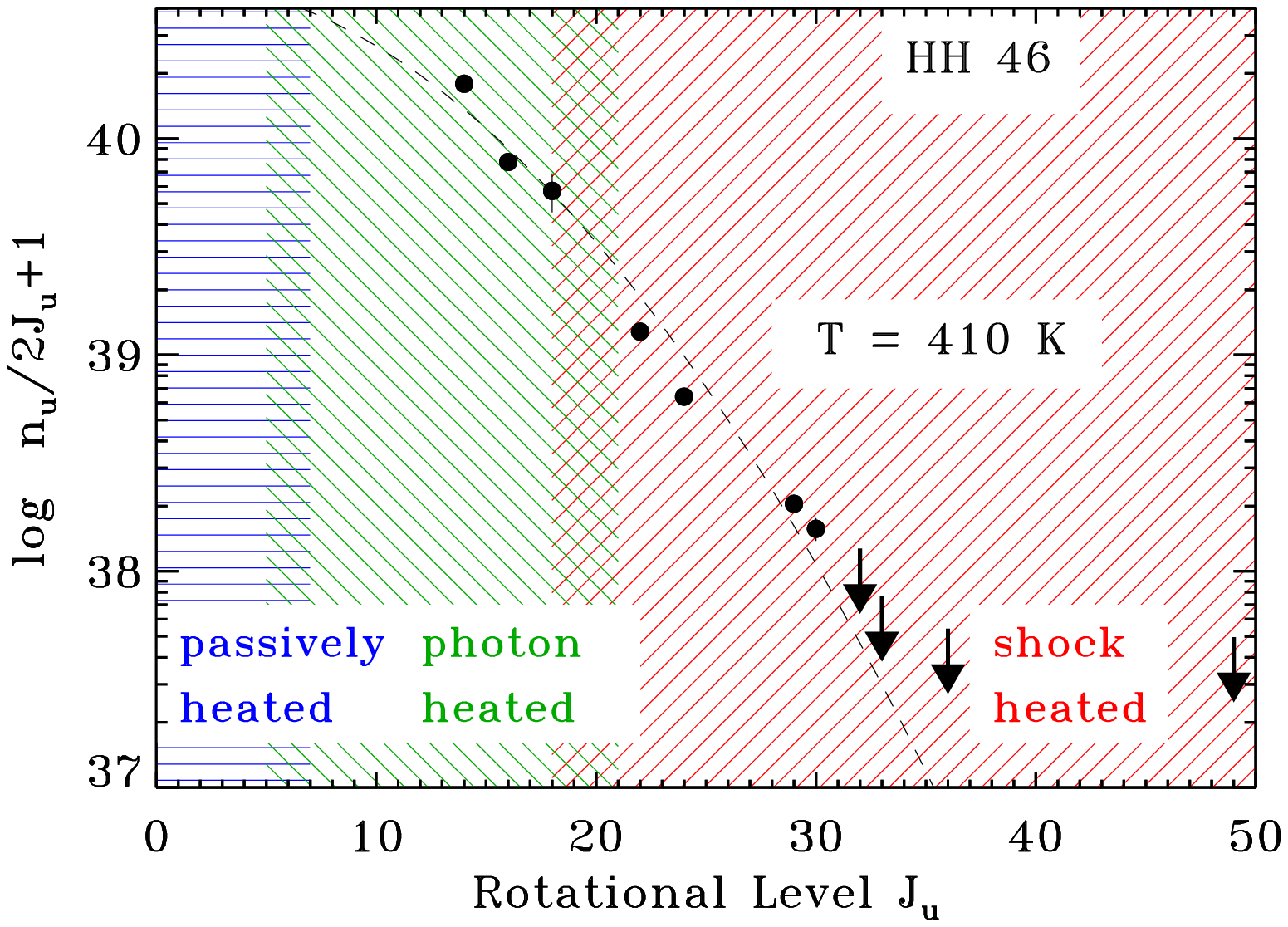}
\end{minipage}\hfill%
\begin{minipage}{0.5\textwidth}
\includegraphics[angle=0,width=\textwidth]{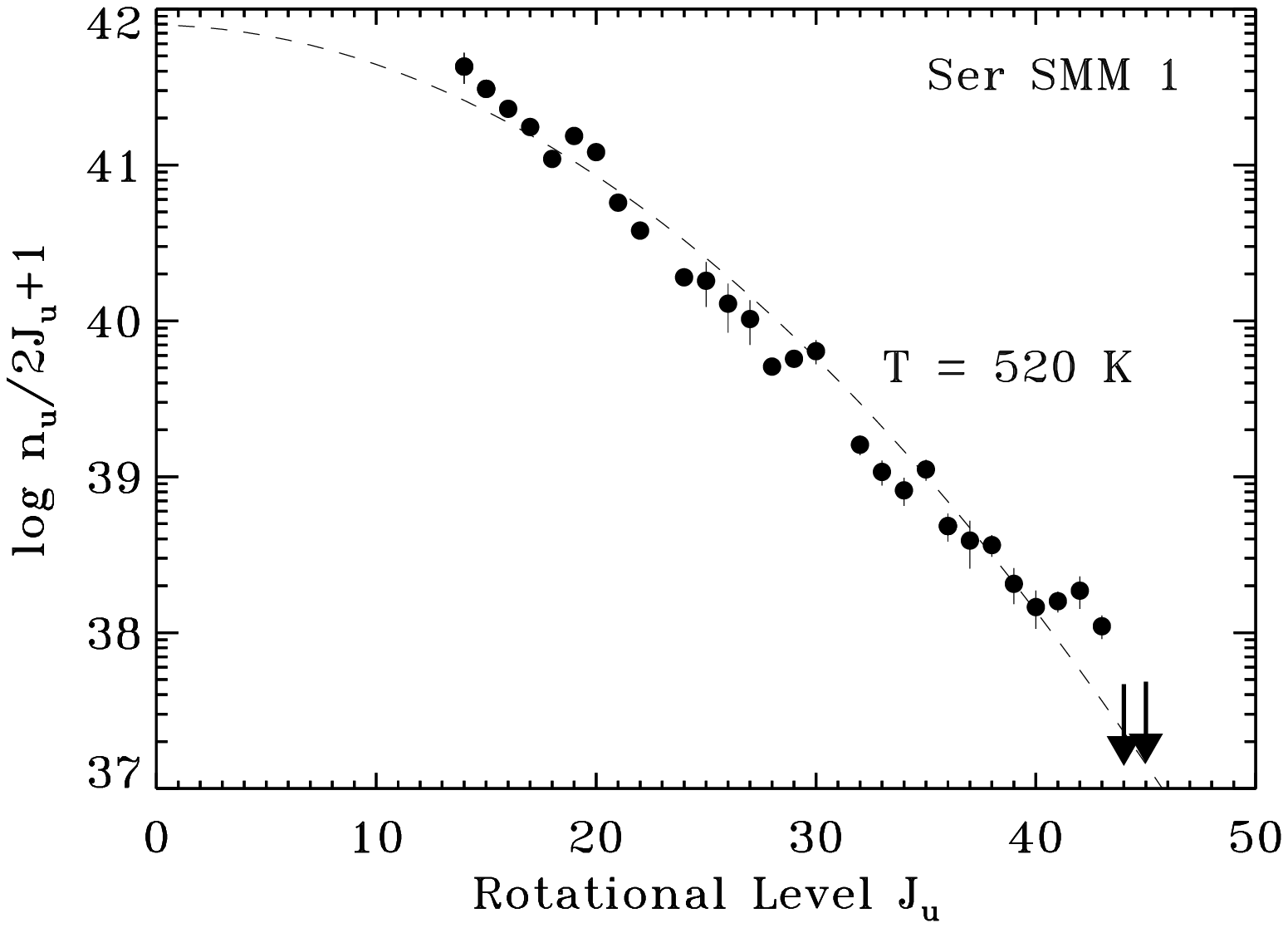}
\end{minipage}\hfill%
\caption{
Diagram illustrating the CO excitation. The number of
molecules in the upper level divided by its statistical weight is
plotted versus rotational quantum number $J_u$.  Left: the low-mass Stage
1 YSO HH 46 (based on \citet{vanKempen10}). Right: the more luminous
embedded source SMM1 (Goicoechea et al.\ in prep.). For the latter
source, a complete PACS spectral scan has been obtained. For both
sources, the fluxes from the central spaxel have been used. The dashed
lines give the thermal distribution at the indicated temperatures. The
hatched areas in the HH 46 diagram indicate the different physical
processes thought to be responsible for the CO emission.
}
\label{fig:coexc}
\end{figure}

The narrow water absorption lines probe the quiescent outer envelope
and any foreground material. Together with the absence of narrow
H$_2^{18}$O $1_{10}-1_{01}$ emission and the maximum narrow emission
that can be hidden in the H$_2$O profiles, these lines constrain the
H$_2$O abundance in the bulk of the cold envelope to $\sim 10^{-8}$.
The lack of (narrow) H$_2^{18}$O $2_{02}-1_{11}$ emission down to
$\sim$ 10 mK rms limits the water abundance in the inner ice
evaporation zone to $< 10^{-5}$ \citep{Visser10}.  In contrast, the
warm quiescent envelope with $T>25$~K is clearly revealed by HIFI in
narrow C$^{18}$O 9--8 emission lines, where the CO abundance jumps by
a factor of at least a few \citep{Yildiz10}.

Figure~\ref{fig:557spectra} includes the $o$-H$_2$O 557 GHz
$1_{10}-1_{01}$ HIFI spectrum of the Stage 1 source TMR-1. This is the
first detection of the H$_2$O 557 GHz line in this Stage.  The fact
that the line is weaker than for the Stage 0 source is consistent with
the non-detection by ISO. Again, the broad line points to an origin in
the outflow shocks rather than the quiescent envelope.

The combination of ground-based, HIFI and PACS data allows the
complete CO ladder from $J$=2--1 up to 43--42 to be characterized for
a low-mass YSO for the first time \citep{vanKempen10}. Because CO is
much less sensitive to chemical effects than H$_2$O, this provides an
opportunity to test various gas heating processes.
Figure~\ref{fig:coexc} shows the CO excitation diagram for the Stage 1
source HH 46, giving a rotational temperature of 410 $\pm$ 25
  K. Also included in this figure are data for another WISH source,
  Serpens SMM1, for which a full PACS spectral scan has been
  obtained. The entire CO ladder can be fit by a single rotational
  temperature of 520 $\pm$ 20 K. CO ladders have also been published
  for two sources from the DIGIT first results, the Stage 1--2
  transitional source DK Cha and the disk around the Herbig star HD
100546 \citep{vanKempen10dkcha,Sturm10}. In both cases, the data
indicate excitation temperatures of several hundred K, but with
evidence for a two temperature component structure in at least one
source. Such a two component fit does not necessarily imply two
different physical regimes or a temperature gradient, however. Optical
depth effects in the CO lines can also lead to curvature in the
rotational diagrams.

Because of the limited information that can be derived from an
excitation diagram, a detailed model has been developed for HH 46
using the known physical structure of the quiescent envelope as a
starting point \citep{vanKempen09hh46}. The passively heated envelope
can explain only the lowest CO rotational transitions, so additional
heating mechanisms are needed (see Fig.~\ref{fig:cartoon}). One
possibility is UV photon heating, in which the UV photons produced in
the immediate protostellar environment can escape through the cavity
carved by the jet and wind from the young star and impinge on the
envelope at much larger distances from the star (up to a few thousand
AU).  It is found that this UV heating of the outflow cavity walls can
reproduce well the intermediate transitions up to $J<20$, depending on
the precise PDR model used. Another mechanism is necessary to populate
the highest CO levels $J>20$: small-scale shocks with velocities $\sim
20$ km s$^{-1}$ pasted along the walls can explain these lines well
\citep{vanKempen10,Visser10}. This model solution is not unique and is
sensitive to model details, but it demonstrates that a good fit can be
found with just two free parameters: the UV flux and the shock
velocity.

Figure~\ref{fig:classIpacs} presents maps of the CO, H$_2$O and [O I]
lines observed with PACS toward the isolated Stage 1 source IRAS
15398-3359, known for its peculiar carbon chemistry \citep{Sakai09}.
All lines are clearly extended in the outflow direction as mapped in
lower-$J$ CO lines \citep{vanKempen09south}, confirming their
association with the shocks and outflow cavities. 

The amount of cooling produced by each of the species can be
quantified from the PACS data. In the central HH 46 spaxel, [O I] and
OH dominate the cooling (at least 60\% of the total cooling in
far-infrared lines) followed by H$_2$O ($\sim 30$\%) and CO ($\sim
10$\%). The total far-infrared line cooling of at least
  $2.4\times 10^{-2}$ L$_\odot$ is consistent with the kinetic
luminosity found by \citet{vanKempen09hh46} from CO outflow maps and
follows the relation found by \citet{Giannini01} between these two
quantities. The [O I] 63 $\mu$m line is spectrally resolved with
PACS, and reveals a jet component with velocities up to 170 km
s$^{-1}$ associated with the blue- and red-shifted outflow seen in
optical atomic lines in less extincted regions of the flow. This is
the first time that such a jet has been seen in a far-infrared line,
providing a powerful diagnostic of the ``hidden jet'' in the densest
optically opaque regions.  OH is detected strongly on source but not
along the outflow axis, in contrast with other species. Both the [O I]
and OH emission are ascribed to a dissociative shock caused by the jet
impacting on the inner dense ($>10^6$ cm$^{-3}$) envelope; disk
  emission is excluded by scaling the observed disk emission from
  Stage 2 sources to the distance of HH 46 \citep{vanKempen10}. In
dissociative shocks, OH is one of the first molecules to reform after
molecular dissociation, whereas the UV radiation from the shock and
the high H/H$_2$ ratio prevent effective H$_2$O formation in the
postshock gas \citep{Neufeld89a}. Once H$_2$ is reformed, the
  heat of H$_2$ formation maintains a temperature plateau in the warm
  postshock gas in which significant H$_2$O is produced
  \citep{Hollenbach89}.

The OH chemistry has been further investigated by \citet{Wampfler10}.
The 163 $\mu$m line seen by PACS toward HH 46 was observed but not
detected with HIFI. The upper limit implies a line width of at least
11 km s$^{-1}$. Inspection of the PACS data for six sources reveals a
correlation between the OH and [O I] fluxes, which, together with the
broad profiles, is consistent with production in (dissociative)
shocks. The relative strengths of the various OH far-infrared lines
are remarkably constant from source to source, indicating similar
excitation conditions. Only the 119 $\mu$m lines connecting with the
ground state vary because they are affected by (foreground)
absorption. Due to the lack of detected quiescent OH (and H$_2$O)
emission in the HIFI spectra, no abundances in the envelope can
yet be determined to test the model predictions of enhanced OH/H$_2$O
ratios due to X-rays.

The HDO/H$_2$O ratio has been determined for the low-mass protostar
NGC 1333 IRAS2A by \citet{Liu10}. Five HDO lines arising from levels
ranging from 20 K up to 170 K have been observed using ground-based
telescopes.  The inner ($>100$ K) and outer ($<100$ K) HDO abundances
with respect to H$_2$ are well constrained to (0.7--1.0)
$\times$ 10$^{-7}$ (3$\sigma$) and (0.1--1.8) $\times$ 10$^{-9}$
(3$\sigma$) respectively.  If the H$_2$O and isotopologue lines
discussed above can be associated with the same component as seen in
HDO, the inferred limit on the warm H$_2$O abundance implies a
HDO/H$_2$O abundance higher than $\sim$1\% in the inner envelope,
consistent with the value found in the hot corino of IRAS16293
($\sim$3\%, \citealt{Parise05hdo}). Although absolute abundances of
HDO and H$_2$O jump by more than one order of magnitude between the
cold and warm envelopes, the HDO/H$_2$O ratio in the outer cold
envelope is found to have a similar value to that of the inner part,
$\sim 1-18$\%.

The detection of HDO lines in other low-mass protostellar sources has
proven difficult from the ground, even under the best weather
conditions. Deep {\it Herschel} observations of high-frequency HDO
lines coupled with deep H$_2^{18}$O data are likely needed to pin down
the cold HDO/H$_2$O ratios.  Alternatively, interferometric
observations of the higher excitation HDO lines, coupled with similar
data of the H$_2^{18}$O $3_{13}-2_{20}$ 203 GHz line (the only
H$_2^{18}$O line readily imaged from the ground,
\citealt{Jorgensen10}), can constrain the HDO/H$_2$O ratio in the warm
gas in the innermost regions. Initial results by \citet{Jorgensen10hdo}
for the NGC 1333 IRAS4B system indicate much lower
values of $< 6\times 10^{-4}$ than those cited above for IRAS2A. This
large difference is not yet understood.

Overall, it appears that the answer to the first question is that
shocks dominate the observed water emission in low-mass YSOs rather
than the hot core.  Both ice evaporation and high temperature
chemistry are likely important in producing high water abundances in
shocks. The answers to the other two questions await more data and
analysis. 

\begin{figure}
\includegraphics[angle=0,width=0.6\textwidth]{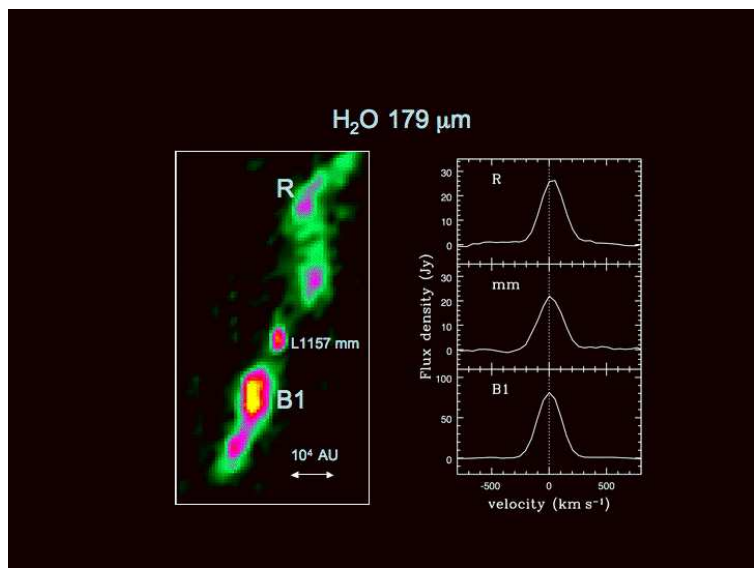}
\caption{Herschel-PACS image of the H$_2$O 179 $\mu$m line toward the
  Stage 0 protostar L 1157, together with (spectrally unresolved) PACS
  spectra at three positions. The water emission is most significant
  in `hot spots' along the outflow where the (rotating) jet interacts
  directly with the cloud as well as toward the protostar itself
  (based on \citealt{Nisini10}). }
\label{fig:L1157}
\end{figure}

\subsection{Outflows from low-mass young stellar objects}
\label{sect:outflow}

{\it Motivation.} Deeply embedded protostars drive powerful molecular
outflows. These outflows extend over arcminute scales for nearby
low-mass objects and can be resolved spatially with {\it
  Herschel}. The brightest outflow knots are clearly separated, often
by arcminutes, from the dense protostellar envelopes targeted in the
low-mass subprogram (\S \ref{sect:lowmass}).  Outflows exhibit a great
diversity in their velocity, degree of collimation, opening angle,
extent, and gas kinetic temperature \citep{Bachiller99}, raising the
question of whether these characteristics can be linked to the
evolutionary state of the driving sources or whether other parameters
(e.g., structure of the protostellar envelope, environment) play a
role. In contrast with ground-based observations of low$-J$ CO
transitions, {\it Herschel} data probe the high-excitation
regions of the flow, where the `hidden jet' interacts with surrounding
material, and they directly target one of the main coolants, H$_2$O,
thus addressing the shock physics and energetics as a function of
distance from the central driving source \citep{Nisini00}.

Outflows represent a major perturbation in the chemistry of the core
gas.  In the youngest systems, the SiO, CH$_3$OH and H$_2$O abundances
are increased by orders of magnitude as a result of grain destruction
and high-temperature gas-phase chemistry.  By determining the
H$_2$O abundances in several spatially and kinematically resolved
components within a given outflow with HIFI, and by doing this for a
sequence of outflows of different ages, the H$_2$O cycle --from grain
mantle, to gas, to mantle-- can be reconstructed.  This allows a
specific puzzle posed by ISO-LWS to be addressed: in most protostellar
outflow regions, the H$_2$O abundance is an order of magnitude lower
than the value of $10^{-4}$ expected if high-temperature reactions
drive all available gas-phase oxygen into H$_2$O or if ice mantles
evaporate.  Ionizing and photodissociative effects of radiation
emitted by nearby dissociative shocks may play a critical role in
explaining the lower-than-expected shocked H$_2$O abundances \citep[see also discussion in][]{Keene97,Snell05}.

Key questions to be addressed in this subprogram are:
(i) What is the nature of the flow: dense molecular jets
or shocked ambient gas along the cavity walls?
Do the characteristics change with distance to the central
driving source and with evolution of the system?
What is the contribution of water emission to the energy
lost in outflows?
(ii) What is the physical impact of the outflows on their surroundings? 
The transfer of energy and momentum from the flow to the envelope and cloud
plays a critical role in controlling future star
formation. 
(iii) What is the chemical effect of the outflows?
Can shock-generated UV radiation modify the physics and chemistry
of nearby molecular gas?  Why is not all oxygen driven into H$_2$O?

{\it Source and line selection.} The outflow subprogram consists of
three parts: a survey part, a line excitation part and spatial maps.
In the survey part, 26 flows from Stage 0 and Stage I low-mass YSOs
are observed in two H$_2$O lines ---the ground-state $o-$H$_2$O 557
GHz line with HIFI down to $\sim$20 mK and the 179 $\mu$m $p-$H$_2$O
line with PACS to similar limits--- in two `hot spots' (blue and
red-shifted outflow lobes), providing a statistical basis for the
assessment of time evolution effects. The driving sources themselves
are covered in the low- and intermediate-mass YSO subprograms (see \S
\ref{sect:lowmass} and \ref{sect:intmass}). The line excitation survey
targets a sub-sample of three Stage 0 outflows for which H$_2$O
emission has previously been detected (L1157, L1448, VLA1623). These
sources are observed in 15 spectral lines with HIFI and PACS (H$_2$O,
OH, [O I] and CO) to characterize their physical structure and energy
budget. Integration times are such that typically 20--40 mK rms is
reached.   In the third part, these same
three flows are spatially mapped with HIFI down to $\sim$20 mK using
the on-the-fly (OTF) mode in the H$_2$O 557 GHz line to determine the
spatial distribution of shocked H$_2$O and compare it with other
outflow tracers like CO, SiO and CH$_3$OH. In addition, they are
imaged with PACS in the 179 $\mu$m line.
For NGC 2071, which is known from {\it Spitzer} data to be
particularly rich in H$_2$O emission \citep{Melnick08}, full PACS
spectral scans are performed at a few positions. 

{\it Initial results.} Figure~\ref{fig:L1157} shows the Herschel-PACS
image of H$_2$O in the 179 $\mu$m line toward the Stage 0 source L1157
obtained during the science demonstration phase of {\it Herschel}
\citep{Nisini10}. PACS spectra at selected positions are included,
illustrating the high $S/N$ of the data. The map beautifully reveals
where the shocks deposit energy into the molecular cloud, lighting up
the water emisson along the outflow. The H$_2$O emission is spatially
correlated with that of pure rotational lines of H$_2$ observed with
{\it Spitzer} \citep{Neufeld09}, and corresponds well with the peaks of
other shock-produced molecules such as SiO and NH$_3$ along the
outflow. In contrast with these species, H$_2$O is also
strong at the source position itself.  The analysis of the H$_2$O 179
$\mu$m emission, combined with existing Odin and SWAS data, shows that
water originates in warm compact shocked clumps of few arcsec in size,
where the water abundance is of the order of $10^{-4}$, i.e., close to
that expected from high-temperature chemistry.  
The total H$_2$O cooling amounts to 23\% of the total far-IR energy
released in shocks estimated from the ISO-LWS data \citep{Giannini01}.

The PACS spectra contained in Figure~\ref{fig:L1157} are spectrally
unresolved. 
HIFI spectra of H$_2$O and CO at the B1 position in the blue outflow
lobe (see Fig.~\ref{fig:L1157}) have been presented and analyzed by
\citet[][]{LeFloch10} as part of CHESS, showing that the H$_2$O
abundance increases with shock velocity (see also \S
\ref{sect:lowmass} and \citet{Franklin08}). \citet{Codella10}
propose that this reflects the two mechanisms for producing H$_2$O:
ice evaporation and high temperature chemistry where the O + H$_2$ and
OH + H$_2$ reactions become significant (see \S~\ref{sect:chemistry}
and Fig.~\ref{fig:network}), with the latter reactions becoming
  more important at higher velocities. The presence of NH$_3$
(another grain surface product) seen by \citet{Codella10} only
at lower velocities is consistent with this picture. The WISH
  data can test this proposed
  scenario at many more outflow positions.

In summary, the limited data to date illustrate the potential of water
to probe and image shock physics and chemistry along the outflows.

\subsection{Intermediate-mass young stellar objects}
\label{sect:intmass}

{\it Motivation.} Intermediate-mass YSOs with $\sim 10^2-10^4$
L$_{\odot}$ (stellar masses of a few to 10 M$_{\odot}$) form the
bridge between the low- and high-mass samples. Besides covering a
significant luminosity range between the two well-studied extremes,
intermediate-mass YSOs are interesting in their own right.  
For example,
they are often found in clustered environments, so they can be used as
laboratories for investigating high-mass cluster regions which are
often more confused due to their larger distances.  Intermediate mass
YSOs may be in an interesting regime where heating from the
outside by radiation from neighboring stars is significant compared with
internal heating \citep{Jorgensen06ori}. This would result in a more
extended warm zone and a different chemical structure of the envelope
than that of their low- and high-mass counterparts. Also, emerging
intermediate-mass YSOs have associated PDRs or reflection nebulae
indicating that not only shocks but also UV photons play a role in
clearing the surroundings.  Finally, they are the precursors of Herbig
Ae/Be stars, whose protoplanetary disks have been very well studied
\citep[e.g.,][]{Meeus01}, but for which very little is known about
their formation history.

Key questions are therefore: (i) What is the structure of the envelopes
of intermediate-mass YSOs?  Do they have a different temperature and
chemistry structure compared with their low- and high-mass
counterparts?  Does the enhanced UV compared with the low-mass
counterparts play a role? (ii) How do intermediate mass YSOs evolve?
To what extent are the answers to the questions in
\S~\ref{sect:lowmass}, \ref{sect:outflow}, \ref{sect:highmass}
different for these sources?

{\it Source and line selection.} Although intermediate mass YSOs are
more energetic than their low-mass counterparts, they are not bright
enough to observe at the distances of their high-mass
cousins. Surprisingly few nearby ($<400$ pc) candidates are known, so
our team has conducted an active search over the past years to
identify a representative sample to observe with HIFI.  Six
intermediate mass YSOs with distances $<$1 kpc have been selected from
this sample and a significant amount of complementary submillimeter
line and continuum observations have been obtained
\citep[e.g.,][]{Fuente05,Crimier10}. All of these sources are still at
the deeply embedded stage, comparable to Stage 0 in the low-mass
case. The {\it Herschel} line selection follows the general criteria
outlined above and the noise levels are similar to those for the
low-mass YSOs.

{\it Initial results.} Observations of various water lines toward NGC
7129 FIRS2 ($430$ L$_\odot$, 1260 pc) are presented in
\citet{Fich10,Johnstone10} and the two lowest $p-$H$_2$O lines
detected with HIFI are included in Fig.~\ref{fig:paraspectra}. The
line profiles are remarkably similar to those of the low-mass YSO NGC
1333 IRAS2A, showing both a broad component (FWHM $\sim$25 km
s$^{-1}$) due to shocks along the outflow cavity as well as a
medium-broad component (FWHM $\sim$6 km s$^{-1}$) associated with
small-scale shocks in the inner dense envelope. Quantitative analysis
of all the H$_2$O and isotopologue lines observed for this object
result in an outer envelope ($<100$~K) abundance of water of order
$\sim 10^{-7}$, an order of magnitude higher than for the low- and
high-mass YSOs.
Alternatively, both the high-$J$ CO and H$_2$O lines seen with HIFI
and PACS can be analyzed in a single slab model with a temperature of
$\sim 1000$~K and density $\sim 10^7-10^8$ cm$^{-3}$, representative
of a shock in the inner dense envelope \citep{Fich10}. This leads to a
typical H$_2$O/CO abundance ratio of 0.2--0.5.

The results for this first source demonstrate that the observing
strategy for this subprogram is sound and that the data can provide
the quantitative information to answer the above questions, once more
sources have been observed and analyzed.

\subsection{High-mass young stellar objects}
\label{sect:highmass}

{\it Motivation.} High-mass stars ($>10$ M$_{\odot}$, $>10^4$
L$_{\odot}$), though few in number, play a major role in the energy
budget and shaping of the Galactic environment
\citep[e.g.,][]{Cesaroni05}.  These are also the type of regions that
dominate far-infrared observations of starburst galaxies.  Their
formation is not at all well understood due to large distances, short
timescales and heavy extinction. 

Recent results show that the embedded phase of the formation of O and
B stars can be divided empirically into several classes.  The earliest
stage are the massive pre-stellar cores (PSC), which are local
temperature minima and density maxima and which represent the initial
conditions.  Their observational characteristics are large column
densities, low temperatures, and absence of outflow or maser activity.
The next stage are the High-Mass Protostellar Objects (HMPOs), where
the central star is surrounded by a massive envelope with a centrally
peaked temperature and density distribution.  HMPOs show signs of
active star formation through outflows and/or masers.  This subprogram
distinguishes mid-IR-bright and -quiet objects (mIRb / mIRq), with a
boundary of 10~Jy at 12~$\mu$m.  The next evolutionary phase is
represented by Hot Molecular Cores (HMC), with large masses of warm
and dense dust and gas, and high abundances of complex organic
molecules evaporated off dust grains.  HMCs have regions of $T>100$~K
that are $>$0.1~pc in size.  The final stage considered here are
Ultracompact HII regions (UCHII), where large pockets of ionized gas
have developed but which are still confined to the star.  UCHII
sources have free-free flux densities of more than a few mJy (for a
distance of a few kpc) and sizes of 0.01--0.1~pc. As with the previous
category, this criterion may imply a luminosity of $>$10$^5$
L$_{\odot}$.  The embedded phase ends when the ionized gas expands
hydrodynamically and disrupts the parental molecular cloud, producing
a classical HII region. Although the evolutionary path of high-mass
YSOs likely runs through these different categories, this is not yet
as well established as for the case of low-mass YSOs, and the physical
structure and processes dominating each of these stages are not well
characterized. {\it Herschel} observations of water and related
species will provide an important piece of this puzzle.

Abundant H$_2$O has been observed with ISO, SWAS and Odin toward the
Orion and SgrB2 high-mass star-forming regions
\citep[e.g.,][]{Cernicharo06,Polehampton07,Persson07}. Toward other
high-mass YSOs, copious hot gas-phase H$_2$O has been detected in
absorption with ISO-SWS \citep{vanDishoeck96,Boonman03h2o} but
not with ISO-LWS. SWAS data indicate generally weak H$_2$O 557 GHz emission
\citep{Snell00}.
Analysis of these data for selected
high-mass star-forming regions implies sharp gradients in the water
abundance, together with strong excitation changes, but their origin is not
yet well understood 
\citep{Boonman03,vanderTak06}.  
More generally, the role of shocks and the relative importance of
shocks versus passive heating in the inner envelope during the
high-mass evolution, either in isolated sources or in clusters, is
currently unclear. In addition, the relation of these phases to the
H$_2$O maser sources needs further study.  Since individual high-mass
sources are at much larger distances (typically a few kpc) than their
low-mass counterparts and thus spatially unresolved, disentangling the
various processes requires observations of numerous H$_2$O transitions
and high spectral resolution, as obtained within WISH using HIFI and
PACS.

Key questions include: (i) What is the chemistry of the warm gas close
to young high-mass stars in various evolutionary stages, specifically
the distribution of H$_2$O?  (ii) What is the relative importance of
shocks vs.\ UV for the interaction of the stars with their
environment?  (iii) What are the kinematics of the cold and warm gas
close to young high-mass stars?  (iv) What are the effects of
clustered star formation and feedback by protostellar outflows on
high-mass regions?

{\it Source and line selection.} 
The WISH Herschel observations of H$_2$O and related species cover
each of the above phases in the formation of high-mass stars.
Pointed observations with HIFI 
are made of a sample of four massive pre-stellar cores and 19 protostellar
sources.  The pre-stellar cores are taken from \citet{Carey98} and
\citet{Pillai06} and are observed only in the 557 GHz line down to
10 mK rms.  The protostellar list consists of ten HMPOs (five mid-infrared
bright and five quiet), four hot cores and five UC HII regions. The sources are
nearby ($<$few kpc) and isolated, selected from surveys by
\citet{Molinari96,Sridharan02,Wood89,vanderTak00}. The basic H$_2$O
line list is the same as for the low-mass and intermediate-mass YSOs,
but with additional H$_2^{18}$O and H$_2^{17}$O lines added because
the main isotopic lines are very bright and highly optically thick.
Absorption lines of H$_2$O at high frequencies
against the optically thick continuum can uniquely probe the kinematics
of the protostellar environment (see
\S~\ref{sect:excitation}). 
Integration times are such that 50--150 mK rms is reached, depending
on the line.

Full PACS spectral scans of all protostellar sources are taken and
provide an unbiased overview of low- and high-excitation H$_2$O lines
(and thus the shocked $\sim$1000 K gas in the beam) together with OH,
[O I] and CO at the position of the protostar, as well as spatial
information on a 50$''$ scale.  $S/N$ ratios $>$100 on the continuum
are reached so that spectrally unresolved lines can be detected.  To
characterize the effects of clustered star formation and feedback,
including the chemistry and cooling rates of intra-cluster gas, HIFI
mapping of the $p-$H$_2$O 1113 GHz line combined with $^{13}$CO 10--9
plus PACS mapping of $o-$H$_2$O 1670 and 1717 GHz down to 300 mK rms
is planned over a region of $\sim 3'\times 3'$ for six nearby
cluster-forming clouds where several hot cores and UCHII regions are
known to co-exist in a small area.

{\it Initial results.}  Figure~\ref{fig:paraspectra} includes the
lowest two $p-$H$_2$O lines for the high-mass YSO W3 IRS5
\citep{Chavarria10}. The profile of the excited $2_{02}-1_{11}$ line
reveals the same broad and medium-broad components as seen for their
low- and intermediate-mass counterparts, but the profile of the
ground-state line is clearly more complex.  In addition to
self-absorption and absorption of the cold envelope against the
continuum, absorptions due to foreground clouds lying along the line
of sight are seen at various velocities. Moreover, for line-rich
sources, weak emission lines due to species other than H$_2$O are
detected.  
The ability of HIFI to disentangle the different physical components is
demonstrated by initial performance verification data on DR21 (OH)
\citep{vanderTak10}. Using a slab model and assuming an $o/p$ ratio of
3 \citep{Lis10}, the broad component due to the outflow is found to
have an H$_2$O abundance of a few $\times 10^{-6}$. Water is much less
abundant in the foreground cloud where $p-$H$_2$O has an abundance
of $4\times 10^{-9}$. The outer envelope abundance is even lower, a few
$\times 10^{-10}$.

To investigate whether these abundances are typical of high-mass
protostellar sources, \citet{Marseille10} model four additional
high-mass YSOs in the lowest two $p-$H$_2$O lines combined with the
ground-state $p-$H$_2^{18}$O line. Thanks to the high spectral
resolution, the envelope emission and absorption can be cleanly
separated from other physical components. \citet{Chavarria10} analyze
a larger set of lines for W3 IRS5. Since the same modeling philosophy
is used for all six sources, the inferred abundances can be directly
compared. The outer envelope water abundances are low in all cases but
show two orders of magnitude variation from $<5\times 10^{-10}$ to
$4\times 10^{-8}$. The origin for these variations is not yet
understood, since no correlation is found with luminosity or
evolutionary stage within this small sample. For W3 IRS5, a jump in
abundance in the inner envelope at $T\approx 100$~K is inferred from
the isotopologue lines, the only indication so far of water emission
from a hot core.

Broad emission profiles due to outflows associated with the high-mass
protostars are commonly seen in the high-mass spectra, testifying to
the importance of shocks. Moreover, the presence of blue-shifted
absorption suggests expansion of the outer envelope rather than
infall. Finally, the high $S/N$ absorption data reveal cold clouds
surrounding the protostellar envelope at velocities offset by just a
few km s$^{-1}$, illustrating the complexity of the protostellar
environment and the unique ability of HIFI and water to detect this
gas \citep{Marseille10}.

In summary, the available data on the first six sources show that
questions (i) and (iii) can be addressed quantitatively. Questions
(ii) and (iv) require more data to answer.

\begin{figure}
\includegraphics[angle=0,width=0.6\textwidth]{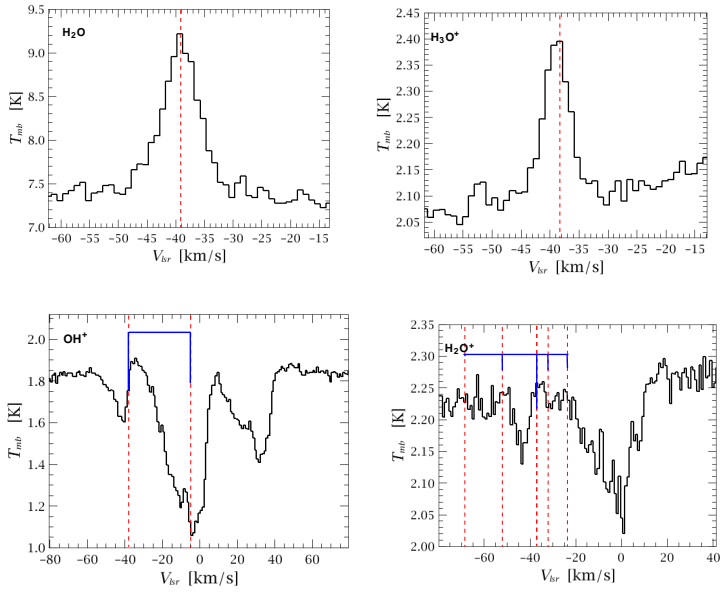}
\caption{Collection of lines of hydrides related to the H$_2$O
  chemistry detected toward the high-mass YSO W3~IRS5 (based on
  \citealt{Benz10}). The dashed red lines indicate the position of the
  main component at the LSR velocity of the protostellar envelope at
  $-38.4$ km s$^{-1}$. For species with fine-structure components
  (OH$^+$, H$_2$O$^+$), the length of the vertical bars indicates the
  theoretical relative intensities. Note that strong OH$^+$ and
  H$_2$O$^+$ absorption occurs around $V_{LSR}\approx 0$ km s$^{-1}$
  in diffuse foreground clouds.}
\label{fig:hydrides}
\end{figure}

\subsection{Radiation diagnostics}

{\it Motivation.}  The H$_2$O chemistry can be strongly affected by
high-energy radiation from YSOs, both UV and X-rays. For typical X-ray
luminosities of $10^{28}-10^{31}$ erg s$^{-1}$ appropriate for
low-mass YSOs, models show that the H$_2$O abundance decreases by two
orders of magnitude or more in the inner envelope, where the X-rays
penetrate up to distances of $\sim$1000 AU (one {\it Herschel} beam;
\citealt{Stauber06}). UV and X-rays also affect the chemistry in
high-mass YSO envelopes, but to a different degree
\citep{Stauber04,Stauber05}. H$_2$O is destroyed not only by direct
photodissociation but also by reactions with ionized molecules, so
that it is important to understand their chemistry in relation to that
of H$_2$O (Fig.~\ref{fig:network}).  Because H$_2$O is not a unique
diagnostic of either UV or X-rays, independent diagnostics need to be
observed. Small hydrides, especially OH$^+$, CH, CH$^+$ and SH$^+$,
are particularly powerful in this regard
\citep{Stauber05,Bruderer10mod}. Recent 2D modeling has shown that
most of the emission from UV tracers originates in the outflow walls,
where their abundances are enhanced by orders of magnitude due to UV
impinging on the walls of the outflow cavities leading to ionization,
dissociation and high-temperature chemistry
\citep{Bruderer09a,Bruderer09b,Bruderer10mod}.  

Key questions are: (i) Can we establish the presence of UV and/or
X-rays through chemistry in deeply embedded objects, where the UV or
X-rays cannot be observed directly? If so, can we quantify it?  (ii)
How do protostellar UV and X-rays influence the surroundings, both the
temperature and the water chemistry?

{\it Source and line selection.} Deep pointed observations with HIFI
are made of lines of seven hydrides that are related to the water
network and sensitive to either FUV or X-rays (or both) in two
high-mass and one low-mass YSO down to 15 mK rms.  Subsequently, a
larger sample of nine low-, intermediate- and high-mass YSOs is
surveyed in the lines of CH, CH$^+$ and OH$^+$. Lines of some hydrides
are also obtained serendipitously with the main H$_2$O settings in the
other subprograms. For example, the CH 536 GHz triplet is often
obtained together with the $o-$H$_2$$^{18}$O $1_{10}-1_{01}$ 548 GHz
line and the H$_2$O$^+$ $1_{11}-0_{00}$ 1115 GHz line with the
$p-$H$_2$O $1_{11}-0_{00}$ 1113 GHz line.  Line frequencies have been
collected, and in some cases first predicted, by \citet{Bruderer06}.

{\it Initial results.} An early surprise from all {\it Herschel-}HIFI KPs is
the detection of widespread H$_2$O$^+$ absorption in a variety of
galactic sources including diffuse clouds, the envelopes of massive
YSOs and outflows
\citep{Gerin10,Ossenkopf10,Benz10,Bruderer10,Wyrowski10,Neufeld10w49,Schilke10,Gupta10}. H$_2$O$^+$
and OH$^+$ lines are even seen in emission in SPIRE-FTS spectra of the
AGN Mkr231 \citep{vanderWerf10}. Model analyses point to an origin in
low density gas ($<10^4$ cm$^{-3}$) with a low molecular fraction
(i.e., a H$_2$/H ratio of a few \%) to prevent H$_2$O$^+$ from
reacting rapidly with H$_2$ to form H$_3$O$^+$. The low molecular
fraction can be the result of either high UV or high X-ray fluxes.

Within WISH, deep integrations on two high-mass YSOs reveal most of
the targeted molecules, a vindication of the model predictions. CH,
NH, OH$^+$ and H$_2$O$^+$ are detected toward both W3 IRS5 and AFGL
2591 \citep{Benz10,Bruderer10}. In addition SH$^+$ and H$_3$O$^+$ are
seen toward W3 and CH$^+$ toward AFGL 2591. Only SH and NH$^+$ are not
detected. H$_2$O$^+$, OH$^+$ and SH$^+$ are new molecules, with OH$^+$
and SH$^+$  only recently detected for the first time in
ground-based observations by \citet{Wyrowski10oh+} and
\citet{Menten11sh+}, respectively.  Figure~\ref{fig:hydrides} contains
a collection of spectra of hydrides most closely related to water.

The spectra show a surprising mix of absorption and emission features.
Three molecules, NH, OH$^+$ and H$_2$O$^+$, appear primarily in
absorption, either in diffuse foreground clouds along the line of
sight, in blue-shifted outflows, or in some cases in clouds close to
the systemic velocity.  For these molecules, the $J=1$ level lies at
47 K or higher.  Limits on the excited state lines give rotational
temperatures $<$13 and $<$19 K for OH$^+$ and H$_2$O$^+$,
respectively.  CH and SH$^+$, for which the $J=1$ line lies at less
than 26 K, are seen in emission, although CH has superposed
absorption. The CH$^+$ $J$=1--0 line occurs mostly in absorption but
the 2--1 line is in emission in AFGL 2591. For W3 IRS5, at least three
H$_3$O$^+$ lines from energy levels lying at a few hundred K above
ground are detected, which combined with ground-based data
\citep{Phillips92} give a rotational temperature of $240\pm 40$~K.
Note that all excitation temperatures are lower limits to the
kinetic temperature of the gas because the lines have high critical
densities ($>10^7$ cm$^{-3}$) resulting in subthermal excitation.

The observed range in line profiles and excitation conditions
immediately indicates that the species originate in different parts of
the YSO environment. The OH$^+$/H$_2$O$^+$ abundance ratio of $> 1$
for the blue-shifted gas in both sources indicates low density and
high UV fields \citep{Gerin10}, as expected along the cavity walls at
large distances from the source. The CH, CH$^+$ and H$_3$O$^+$
emission lines must originate closer to the protostar, at densities
$>10^6$ cm$^{-3}$ \citep{Bruderer10}. Their abundances are roughly
consistent with the 2D photochemical models, in which high temperature
chemistry and high UV fluxes boost their abundances along the outflow
walls \citep{Bruderer09b,Bruderer10mod}. The UV field is enhanced by
at least four orders of magnitude compared with the general interstellar
radiation field in these regions.

Because the H$_2$O$^+$ ground state line occurs close to that of
$p-$H$_2$O, observations of both species are available for a large
sample of high-mass protostars within WISH. Using also H$_2^{18}$O
data, \citet{Wyrowski10} provide column densities for both species for
all components along 10 lines of sight.  H$_2$O$^+$ is always in
absorption, even when H$_2$O is seen in emission.  Consistent with the
above scenario, the highest H$_2$O$^+$ column densities are found for
the outflow components.  Overall, H$_2$O$^+$/H$_2$O ratios range from
0.01 to $>1$, with the lower values found in the dense protostellar
envelopes and the larger values in diffuse foreground clouds and
outflows.

In summary, most of the targeted hydrides are readily detected and
they appear to be even more abundant and widespread than expected
prior to Herschel. When associated with protostars, they trace regions
of high temperatures and enhanced UV fluxes, such as found along
cavity outflow walls. However, the fact that some hydrides are seen in
absorption whereas others appear in emission is not yet fully
understood.

\subsection{Circumstellar disks}
\label{sect:disks}

{\it Motivation.} A fraction of the gas and dust from the protostellar
envelope ends up in a rotating disk around the pre-main sequence
star ($\leq 10$ Myr), thereby providing the basic material from which
planetary systems can be formed.
Recent observations and modeling have shown that many young disks have
a strongly flaring structure in which the surface layers are exposed
to intense ultraviolet radiation from the star. This results in a
highly layered structure with a hot top layer which is mostly atomic
due to rapid photodissociation; an intermediate warm layer with an
active chemistry and responsible for most of the observed molecular
emission; and a cold, dense midplane layer where most species are
frozen out onto grains \citep[e.g.,][]{Aikawa02,Bergin07ppv}.

A major question in disk evolution and planet formation studies is the
amount of grain growth and vertical mixing as a function of radial
position and time. In the static models described above, little
gas-phase H$_2$O is expected in the outer disk, with most of it frozen
out as H$_2$O ice in the midplane. However, if the H$_2$O ice is
regularly circulated to the warmer upper layers, the gas-phase H$_2$O
abundance increases by orders of magnitude \citep{Dominik05}. This in
turn affects the gas temperature and the abundances of other species,
including its photodissociation products OH and O.  Indeed, 
analysis of CO observations shows that significant gaseous CO is
present at temperatures below that for freeze-out \citep[e.g.,][]{Dartois03}.
Both photodesorption of ices and rapid vertical mixing combined with a
steep vertical temperature gradient and grain growth have been invoked
to explain this \citep{Aikawa07,Oberg07,Hersant09}.

HIFI observations are uniquely suited to constrain vertical mixing
models because H$_2$O is the major ice reservoir and its line emission
is particularly sensitive to temperature. HIFI thus probes the cold
H$_2$O gas reservoir in the outer ($>30$ AU) part of disks.  These
observations are therefore highly complementary to {\it Spitzer}
detections of hot H$_2$O lines which originate in the inner few AU of
disks \citep{Salyk08,Carr08}. 
Key questions are: (i) What do H$_2$O observations of circumstellar
disks tell us about grain growth and the importance of vertical mixing
from the cold midplane to the warm surface layers where ices 
photodesorb or thermally vaporate? (ii) What is the relative importance of
photodissociation and desorption? (iii) What are the implications of
the {\it Herschel} observations of the cold water reservoir in disks
for the transport of water to the inner planet-forming zones of disks?

{\it Source and line selection.} Deep integrations down to a $\sim$1.5
mK rms in 0.5 km s$^{-1}$ bin are performed for the $o-$H$_2$O 557 GHz
line in four targets (TW Hya, DM Tau, LkCa15 and MWC 480) and down to
$\sim$4 mK rms for the $p-$H$_2$O 1113 GHz line in two targets (TW Hya
and DM Tau). The primary aim of these very deep integrations is to put
stringent limits on the {\it cold} water reservoir. Shorter
integrations down to 10 mK rms in the 557 GHz line are performed for eight
additional targets to determine whether the four deep targets are unusual
in their H$_2$O emission.  The excited H$_2$O $3_{12}-3_{03}$ 1097 GHz
and CO $J$=10-9 line are observed in two sources to probe the warmer
H$_2$O reservoir with HIFI. Sources have been selected to be clean
from surrounding cloud emission in single dish observations
\citep[cf.][]{Thi01}.

{\it Initial results.} Deep integrations on the DM Tau disk show a
tentative detection of the $o-$H$_2$O line with a peak temperature of
$T_{\rm MB}$=2.7 mK and a width of 5.6 km s$^{-1}$
\citep{Bergin10}. Regardless of the reality of this detection and the
uncertainty in the collisional rate coefficients, the inferred column
density is a factor of 20--130 weaker than predicted by a simple model
in which gaseous water is produced by photodesorption of icy grains in
the UV illuminated regions of the disk.  The most plausible
implication of this lack of water vapor is that more than 95-–99\% of
the water ice is locked up in large coagulated grains that have
settled to the midplane and are not cycled back up to the higher disk
layers. Thus, the deep HIFI data can directly address
the key questions of this subprogram, even without actual detections.

\section{Discussion}

Although less than 10\% of the WISH data have been analyzed to date,
several general trends are emerging.

\subsection{Trends with evolutionary stage}

The development of the spectral features with evolutionary phase is
illustrated by Fig.~\ref{fig:557spectra} for low-mass sources. Water
emission is not detected from cold dense cores prior to star
formation, down to very low levels. The implication is that most of
the water is frozen out as ice on dust grains in the densest regions
with at most a small layer of water vapor due to photodesorption of
ice in the outer regions of the core.  As soon as the protostar turns
on in the center, however, prominent H$_2$O lines appear. One of the
main surprises for the low mass YSOs is that this emission has broad
profiles, even for H$_2$$^{18}$O, indicating an origin in shocked
gas. This association with outflows is directly revealed by imaging of
the water line (Fig.~\ref{fig:L1157}). H$_2$O abundances in the
outflow are high, of order $10^{-5}-10^{-4}$ (assuming a CO/H$_2$
abundance ratio of $10^{-4}$), increasing with higher velocity. 
In contrast, quiescent envelope abundances, as derived from narrow
absorption features and the lack of narrow H$_2$$^{18}$O emission, are
low, of order $10^{-8}$. 

The last evolutionary stage of protoplanetary disks when the envelope
has dissipated shows again little H$_2$O emission down to very low
levels (Fig.~\ref{fig:557spectra}).  This is the first time that the
cold water reservoir in the outer regions of disks has been
probed. The lack of emission implies that the upper layers of this
disk are surprisingly `dry', both in the gas and in the solid
phase. About 95--99\% of the water ice appears to be contained in
large icy grains in the disk midplane which are not circled back up to
higher layers. If confirmed by observations of additional sources,
this may have profound implications for our understanding of the
chemistry and physics of the outer disks. On the other hand, lines of
warm and hot water have been detected in other disks with {\it
  Spitzer} \citep{Pontoppidan10} and with PACS \citep{Sturm10},
  although the latter detection needs confirmation.. Tying these data
together should provide a nearly complete view of the distribution of
water vapor from the inner to the outer disk.

For all of these stages, the availability of data on both $o-$H$_2$O
and $p-$H$_2$O indicates that the low water abundance is robust and
independent of any $o/p$ ratio of water. The overall emerging
picture is therefore that most water is formed on grains in the
molecular cloud and pre-stellar core phases, is present mostly as ice
in the quiescent envelope and is likely transported in this form into
the outer part of young disks without further evaporation,
consistent with evolutionary models by \citet{Visser09}.

\subsection{Trends with luminosity}

Fig.~\ref{fig:paraspectra} illustrates the trend with luminosity for
the embedded protostellar evolution stage. An initial conclusion is that
the line profiles are remarkably similar in spite of the four orders of
magnitude range in luminosity. All profiles show a broad (FWHM 25--40
km s$^{-1}$) component due to the shocks along the outflow cavity
walls on scales $>1000$ AU (`shell shocks'), and a medium-broad (FWHM
5--10 km s$^{-1}$) component ascribed to shocks caused by the
interaction of jets and winds with the inner dense ($>10^6$ cm$^{-3}$)
envelope on scales $<1000$ AU. Thus, the processes by which the jets
and winds that are launched close to the star interact with their
environment are apparently similar across a wide range of masses and
luminosities.  The H$_2$O medium-broad profiles clearly have larger
line widths than those of other high density envelope tracers such as
HCO$^+$ 4--3 and CS 7--6, but generally agree with the profiles of other
grain surface products such as CH$_3$OH.

For low- and intermediate mass YSOs, the outflows dominate the H$_2$O
emission whereas for high-mass YSOs, narrower lines from the passively
heated envelope become more prominent, especially for H$_2$$^{18}$O
and H$_2$$^{17}$O. Inferred outer envelope abundances based on the
isotopologue data and narrow absorption features vary between
$10^{-8}$ and $10^{-10}$ with no obvious trends across luminosity or
evolutionary stage, except for a potentially higher value for one
intermediate-mass source. These low abundances thus confirm the
earlier SWAS conclusions but can significantly refine them because of
the availability of more lines from both $o-$ and $p-$H$_2$O allowing
more accurate determinations. Because of the dominance of outflow
emission, it is more difficult than expected to put firm constraints
on the inner quiescent `hot core' water abundances. Generally the
`stirred-up' inner region probed by the medium-broad component
overwhelms any hot core emission, even though this warm quiescent gas
is clearly seen in, for example, C$^{18}$O $J$=9--8 or 10--9 emission.
Simultaneous analysis of these CO high$-J$ data with the H$_2$O lines
may help to further disentangle the components. Indeed, the coverage
of the entire CO ladder up to $J=35$ with all three instruments on
{\it Herschel} will soon become one of the most important diagnostics
to characterize the physical processes in a wide variety of sources.

The PACS and HIFI observations of high-$J$ CO, H$_2$O and [O I]
illustrate the need for the development of multi-dimensional models
which include UV photon and shock heating along the cavity walls for
both low- and high-mass sources.  This need is strengthened by the
detection of various hydrides enhanced by UV radiation, including
OH$^+$, H$_2$O$^+$, SH$^+$, CH and CH$^+$. Initial quantitative
estimates of the total far-infrared line emission indicate that H$_2$O
is not the major coolant of the gas, even in the densest ($>10^6$
cm$^{-3}$) regions, but this needs to be confirmed by
observational data at more positions, both on and off source. [O I]
and OH have higher luminosities and appear to trace a dissociative
shock in the densest inner part.

\subsection{Water chemistry}

The {\it Herschel} data demonstrate that all three chemical networks
illustrated in Fig.~\ref{fig:network} are important in star-forming
regions. Solid-state chemistry dominates in the coldest regions, as
evidenced by the absence of H$_2$O line emission from cold pre-stellar
cores and the low gaseous water abundances in the outer protostellar
envelopes. Most of the water is in the form of ice formed through the
reactions illustrated in the bottom part of Fig.~\ref{fig:network}.
The tiny amounts of water gas detected with HIFI either result from
non-thermal desorption of water ice or from gas-phase ion-molecule
reactions, or from a combination of both.  If the absorption of H$_2$O
gas toward the pre-stellar core L1544 is confirmed, this first direct
detection of water vapor in a quiescent dark cloud will allow a
quantitative test of these chemical models. Similarly, a thorough
analysis of the abundance structure of H$_2$O gas and ice for those
protostellar envelopes where both species are detected can test key aspects of
the cold chemistry. Eventually, fully coupled time-dependent gas-grain
chemistry models should include the new solid-state laboratory data on
water ice formation. Such models can then also be used to interpret
the puzzling HDO/H$_2$O abundance ratios found in different regions.

The detection of {\it all} the hydrides involved in the ion-molecule
scheme leading to water ---OH$^+$, H$_2$O$^+$, H$_3$O$^+$, HCO$^+$ and
OH--- will be one of the lasting legacies of the {\it Herschel}
mission. While these species have been part of the reaction networks
for decades, the ease with which rapidly reacting ions such as OH$^+$
and H$_2$O$^+$ are detected has surprised even the
astrochemists. Since all key reaction rates involved are well
understood, the abundance ratios can be used to probe the physics of
the regions in which they are found, such as high UV fields and low
densities.  The final clue to the basic oxygen gas-phase chemistry
network will be provided by deep searches for O$_2$, which are planned
for some of the WISH sources in other key programs.

The bright water emission from protostars is dominated by $C-$type
shocks where the temperature rises to a few thousand K. The
high-quality line profiles may allow the two chemical regimes to be
separated: sputtering of icy mantles and high-temperature
chemistry. Whether all available oxygen is driven into water at the
highest temperatures and velocities is still an open
question. OH/H$_2$O ratios derived from velocity-resolved OH line
profiles could form a powerful test of the neutral-neutral reactions.
If the contribution from dissociative $J-$type shocks can be cleanly
separated, as appears to be the case for HH 46, the observed O/OH
abundance ratios would provide a further test of the high temperature
gas-phase reaction networks, but now under conditions with much higher
H/H$_2$ ratios in the presence of UV radiation.

Finally, all three chemical networks play a role in the water
chemistry in disks, with high temperature chemistry dominating in the
inner disk and ice chemistry in the outer disk.  Here {\it Herschel}
teaches us another lesson: the importance of coupling the water
chemistry with the dynamics and physical evolution of the disks, in
particular the growth and settling of the icy grains to the midplane.

\section{Conclusions}

The goal of the WISH program is to use water and related species as
physical and chemical probes of star-forming regions over a range of
luminosities and evolutionary stages. The initial results presented in
\S 4 and discussed in \S 5 demonstrate that many of the questions
asked in the individual subprograms can indeed be addressed by the
WISH data, thanks to the combination of excellent sensitivity, fully
resolved line profiles and spatial information, thereby validating the
observational strategy.  Initial surprises include a near absence
  of gaseous water in pre-stellar cores and disks, the dominance of
shocks rather than hot cores in controlling the bright water emission
from protostars, and the detection of all of the ions and hydrides
involved in the water chemistry schemes. For pre-stellar cores
  our observations are in agreement with models that suggest near
  total freeze-out of water in the form of ice, whereas in disks the
  inferred water abundance is at levels significantly lower than
  predicted.  So far, water has not yet been found to be the
major coolant in any of these regions. All remaining observations are
expected to be taken within the next year. Quantitative analysis will
require the further development of multidimensional models of
protostellar envelopes, outflows and disks. Together with the results
from related {\it Herschel} key programs, they will greatly enhance
our understanding of water in the galactic interstellar medium and
solar system, and provide a true legacy to follow the water trail from
the most diffuse gas to dense cores and disks, and eventually comets
and planets in our own Solar system.

\section{Acknowledgments}

The authors are grateful to the HIFI project scientist, Xander
Tielens, for helping to ensure the powerful HIFI capabilities for
water observations, and for his encouragements over the last
decade. They salute the HIFI and PACS instrument builders for
providing two superb scientific instruments, and are much indebted to
the laboratory and theoretical chemistry groups for providing the
necessary molecular data to analyze and interpret the water data.
They thank the referee, David Hollenbach, for his constructive
comments on the manuscript and Laurent Wiesenfeld for helpful
discussions on the H$_2$O-H$_2$ collisional rate coefficients.  We
thank the Netherlands Organization for Scientific Research (NWO) and
many other national funding agencies for their financial support.

HIFI has been designed and built by a consortium of institutes and
university departments from across Europe, Canada and the United
States under the leadership of SRON Netherlands Institute for Space
Research, Groningen, The Netherlands and with major contributions from
Germany, France and the US. Consortium members are: Canada: CSA,
U.Waterloo; France: CESR, LAB, LERMA, IRAM; Germany: KOSMA, MPIfR,
MPS; Ireland, NUI Maynooth; Italy: ASI, IFSI-INAF, Osservatorio
Astrofisico di Arcetri- INAF; Netherlands: SRON, TUD; Poland: CAMK,
CBK; Spain: Observatorio Astron\'omico Nacional (IGN), Centro de
Astrobiolog{\'{\i}}a (CSIC-INTA). Sweden: Chalmers University of
Technology-MC2, RSS \& GARD; Onsala Space Observatory; Swedish
National Space Board, Stockholm University - Stockholm Observatory;
Switzerland: ETH Zurich, FHNW; USA: Caltech, JPL, NHSC.

HCSS / HSpot / HIPE are joint developments by the Herschel Science
Ground Segment Consortium, consisting of ESA, the NASA Herschel
Science Center, and the HIFI, PACS and SPIRE consortia.

\newpage







\bibliographystyle{apj}
\bibliography{biblio_evd}


\end{document}